\documentclass[manuscript]{acmart}

\usepackage{enumitem}
\usepackage{amsfonts} 
\usepackage{amsmath}
\usepackage{natbib}
\usepackage{cleveref}
\usepackage[linesnumbered,ruled,vlined]{algorithm2e}
\usepackage {algpseudocode}
\usepackage{algorithmicx}
\usepackage{algcompatible}
\usepackage{booktabs}
\usepackage{multirow}
\usepackage{algpseudocode}
\usepackage{graphicx}
\usepackage{subcaption}
\usepackage{cleveref}

\usepackage{longtable}

\usepackage{soul}
\usepackage{xcolor}
\usepackage{colortbl}
\usepackage{nicefrac}

\newcommand{\ind}[1]{\mathbb{I}\left\{ #1 \right \}}
\newcommand{\participant}[1]{{\textit{``#1''}}}

\AtBeginDocument{%
  }

\setcopyright{acmlicensed}
\copyrightyear{2024}
\acmYear{2024}
\acmDOI{XXXXXXX.XXXXXXX}

\acmConference[]{}{}{}

\acmSubmissionID{2489}

\begin{document}

\title{A Sim2Real Approach for Identifying Task-Relevant Properties in Interpretable Machine Learning}

\author{Eura Nofshin}
\authornote{Both authors contributed equally to this research.}
\email{eurashin@g.harvard.edu}
\orcid{}

\author{Esther Brown}
\authornotemark[1]
\email{estherbrown@g.harvard.edu}

\author{Weiwei Pan}
\email{weiweipan@g.harvard.edu}

\affiliation{%
  \institution{Harvard University}
  \city{Cambridge}
  \state{MA}
  \country{USA}
}

\author{Brian Y. Lim}
\email{brianlim@comp.nus.edu.sg}
\affiliation{%
  \institution{National University of Singapore}
  \city{Singapore}
  \country{Singapore}
}

\author{Finale Doshi-Velez}
\email{finale@seas.harvard.edu}
\affiliation{%
  \institution{Harvard University}
  \city{Cambridge}
  \state{MA}
  \country{USA}
}

\renewcommand{\shortauthors}{Nofshin, Brown, Pan, Lim, Doshi-Velez}

\begin{abstract}
Explanations of an AI's function can assist human decision-makers, but the most useful explanation depends on the decision’s context, referred to as the downstream \textit{task}. User studies are necessary to determine the best explanations for each task. Unfortunately, testing every explanation and task combination is impractical-- especially considering the many factors influencing human+AI collaboration beyond the explanation's content.

This work leverages two insights to streamline finding the most effective explanation. First, explanations can be characterized by \textit{properties}, such as faithfulness or complexity, which indicate if they contain the right information for the task. Second, we introduce XAIsim2real, a pipeline for running synthetic user studies. In our validation study, XAIsim2real accurately predicts user preferences across three tasks, making it a valuable tool for refining explanation choices before full studies. Additionally, it uncovers nuanced relationships, like how cognitive budget limits a user's engagement with complex explanations--  a trend confirmed with real users.

\end{abstract}



\received{12 September 2024}
\received[revised]{12 March 2009}
\received[accepted]{5 June 2009}

\maketitle

\section{Introduction}
\label{sec: introduction}
AI decision support that also provides explanations can help people perform their tasks better.  For example, suppose a bank is using an AI decision support to assist in making loan decisions.  Information about if and how the AI used protected features, such as gender, when vetting a particular candidate could assist the human decision-maker in ascertaining whether it is appropriate to use the AI recommendation.  Here, the user's \textit{task} is determining whether the AI used protected features inappropriately---and the explanation of the AI provides insight into the AI's reasoning to enable this adjudication.  

However, the reality is that explanations do not always help performance in human+AI decision making \citep{rong2023surveyUserStudies, bansal2021AITeamPerformance, chen2023understandingOverreliance}, and some studies have even found that explanations worsened user performance by increasing their over-reliance on the AI \citep{bansal2021AITeamPerformance, chen2023understandingOverreliance}.  Many design factors impact the usefulness of the explanation in these studies, ranging from presentation \citep{lei2024UI} and the user's level of expertise \citep{morrison2024impact} to cognitive biases \citep{nourani2021anchoring} and the user's reasoning patterns \citep{wang2019designing}.

In this paper, we focus on one particular design aspect: the need for the  \textit{content} of the explanation to be appropriate for the user's task \citet{lai2023selective}.  In our loan example above, it would not be helpful if the explanation said that the model rarely uses gender in its recommendations; the loan officer needs to know if and how the model used gender in this specific decision.  Recent years have seen an explosion in the number of post-hoc explanation methods that can be used to provide insight into a model's reasoning~\citep[e.g.][]{ribeiro2016LIME, lundberg2017SHAP, smilkov2017smoothgrad, sundararajan2017IntegratedGradients}.  Alongside these explanation methods, have come user studies to test which methods perform best on a specific task~\citep[e.g.][]{hase2020evaluating, jeyakumar2020can}. 

However, our current approach to user studies, in which a fixed set of explanation methods (e.g. LIME, SHAP, Anchor, etc.) are tested to determine which most improves the user's task performance, does not produce \textit{generalizable} knowledge that can be used to select explanations for a future task. At the end of a study, we do not know \textit{why} one explanation (such as LIME) worked better than another (such as Anchor) for a task.  When faced with a new task, one has to do another user study with a large number of explanation methods to choose from.

We argue that the first step towards more effectively and efficiently identifying the appropriate explanation content for a new task is to consider the explanation's \textit{properties}.  Rather than thinking in the design space of specific methods for generating explanations (e.g. is LIME better than Anchor?), properties get us closer to the mechanism by which the explanation may provide value.  For example, in the loan task above, any \textit{faithful} explanation---one that exposes all the features used in a decision---may be sufficient to help the user realize that the model is using a protected feature in this specific decision. Indeed, the post user-study discussion in many works \citep[e.g.][]{hase2020evaluating, jesus2021can} posit various explanation properties as the mechanism for their results. Once we identify which properties are required for a task, we can easily identify explanations that satisfy those properties. However, current works have yet to directly use properties to limit the design space in identifying what explanations may be useful for what task. 

It is not sufficient to move directly from the design space of explanation methods to explanation properties. Tens of properties have been proposed in the ML literature without validation that they are helpful to real users ~\citep{lu2021crowdsourcing, nauta2023anecdotalQuantitative}. Complicating matters further, each intuitive property of explanations, like "faithfulness," can have multiple, sometimes conflicting, mathematical formalizations \citep[e.g.][]{yeh2019fidelity, barr2023disagreementFaithfulness, dasgupta2022frameworkEvaluatingFaithfulness}.  A recent paper identified over 100 mathematical formulations across different properties, with over 23 different mathematical formalizations for robustness alone~\citep{chen2022harmonized}). While the design space of properties brings us closer to what makes an explanation's content useful or not, it is still impractical to naively run user studies in hope of identifying which properties are most relevant for a new task. 


This observation brings us to our contribution.  In this work, we introduce XAIsim2real, a framework that: (1) uses a computational pipeline to run \textit{simulated} user studies, and (2) validates the most promising explanations from those simulations in \textit{real} user studies. While a few prior works have also proposed simulated user studies~\citep{chen2022useCaseSimulations}, they do not use our first insight about the importance of explanation properties as the design space.  Rather than test different explanation methods, our computational framework directly optimizes explanations for different properties and tests, in-silico, whether they are the most promising ones to carry into studies with real users.

To demonstrate our XAIsim2real framework, we perform a proof-of-concept study wherein we first use our computational pipeline to generate hypotheses, then design and run real-user studies to validate these hypotheses. Our validation study explores the following research questions: 
\begin{itemize}
        \item [\textbf{RQ1.}] How well does the performance of simulated users match that of real users?
        \item [\textbf{RQ2.}] Can we formally link different explanation properties to user performance on different tasks? 
        \item [\textbf{RQ3.}] How do our assumptions about humans, functions, tasks, and explanations affect the relationship between properties and task-performance? 
\end{itemize}

Specifically, we use our XAIsim2real framework to form and test hypotheses on which explanation properties will be most helpful for users in three common tasks: forbidden features (check whether AI's decision relied on an inappropriate feature), counterfactual simulation (answer ``what if'' questions about the AI's decision if its input were to change), and forward simulation (predict the AI's decision). 
Our in-silico simulations were used to hypothesize which explanation properties would be important for what tasks, and then we ran a user study to test whether those hypothesized relationships held for real users. In line with recent literature that suggests different properties are needed for different tasks \citep{liao2022connectingUsageContexts, lai2023selective, chen2023understanding, jesus2021can, wang2019designing, liao2020questioning, lim2009assessing}, we found that faithfulness is important for forbidden features and forward simulation, whereas robustness is important for counterfactual simulation. Moreover, our in-silico studies hypothesized that the importance of explanation sparsity depended the user's cognitive budget.  \textit{XAIsim2real correctly anticipated all of these findings before we ran our human user studies, suggesting that our framework can be used to effectively identify promising candidates in advance of running expensive user studies.}

Our demonstration study suggests that our XAIsim2real framework has promise for a wide variety of applications.  It enables \textit{practitioners} to identify the most promising set of explanations to compare in a real user study. It also empowers \textit{researchers} to systematically explore connections between explanation properties and user task-performance in-silico, and then validate the connections in real user studies.  Importantly, the practitioner and researcher can explicitly vary the type of user they expect to see in the study, allowing them to explore the implications of different factors affecting human decision-making.  For example, we explored the impact of a limited cognitive budget; our framework can also include other factors such as domain expertise or cognitive biases.  These factors could be used to further understand differences in real users: for example, in our demonstration study, we found that our proxy users were most predictive of top-performing user's decisions.  

\section{Related work}
\label{sec: related-work}

\subsection{User studies for evaluating explainable AI (XAI)}
\label{sec: related-work-user-studies}
User studies are important for comparing the relative helpfulness of different explanation methods; a recent survey of user studies in XAI identified $97$ papers from top-venues in the last five years that had a user study component~\citep{rong2023surveyUserStudies}. 
The set of explanations tested in these user studies is typically chosen because they are popular ~\citep[e.g.][]{jesus2021can, hariharan2023xai, karagoz2024evaluating} or this set is crafted specifically for the application domain \citep[e.g.][]{nadeem2023sokSecurity, paredes2021importanceCyber, scarpato2024evaluatingClinicians}. 
This approach for selecting candidate explanations results in generalizability and reproducibility issues. For example, it is unclear what made LIME best in one study, but not in another --  \citet{chen2022useCaseSimulations} found LIME to best for counterfactual simulation, but \citet{hase2020evaluating} found another method-- Prototype-- to be better. 

In XAIsim2real, we select candidate explanations based on their properties, which we believe improves the generalizability of our insights, because properties introduce a layer of abstraction between the explanation and its context \citep{chen2022harmonized}. 

The above studies assume that \textit{any explanation} is better than \textit{no explanation}. 
\citet{bansal2021AITeamPerformance} found that explanations can cause users to over-rely on AI systems, which hinders their performance . In response, there is growing number of studies that seek to mitigate user over-reliance, for example, by changing the point at which an AI's interacts with the user as well as what explanations are given \citep{buccinca2021trust}. 

In our study, we focus on tasks where the explanation is \textit{required} for the user to make an informed decision-- to forward simulate the AI's prediction, the user needs to know what features are important. We choose these tasks so that we can isolate the effect of the explanation property on task-performance. 
That said, our work to understand how properties affect performance can help reduce over-reliance. For example, \citet{lai2023selective} argues that we can reduce over-reliance by tailoring an explanation's content to the user's task-specific needs. Our work on identifying task-relevant properties offers direct insight into how to tailor the explanations to a task. 


In several user studies, authors discussed how properties may have affected performance in a \textit{retrospective} manner. That is, after running the study, they speculated on which properties made the explanations successful (or not). For example, existing works have hypothesized that counterfactual simulation may need faithful explanations \citep{hase2020evaluatingExplainable}, forming good mental models may need faithful explanations that also have low complexity \citep{kulesza2013too}, and joint decision-making may need robust explanations \citep{jesus2021can}.\footnote{In each case, specific definitions of properties (e.g. faithfulness, complexity, robustness) are chosen by the authors; not all definitions are the same.} Rather than looking for the effects of explanation properties after we run user studies, with XAIsim2real, we form hypotheses on the effect that the properties will have on performance beforehand, and design studies to test these hypotheses.

Finally, very few existing studies have aimed to test the effect of a specific property on task-performance: \citet{poursabzi2021manipulating} studied how explanation compactness affects forward simulation, \citet{nguyen2018comparing} studied the same but with faithfulness, and \citet{lage2019evaluation} studied how types of complexity (explanation size, cognitive chunks) affect several tasks (forward simulation, counterfactual simulation, verification). However, in these examples, the property to be studied was selected based on prior work or domain knowledge and does not provide a general method for generating candidate properties for user studies. Taking a different approach, \citet{liao2022connectingUsageContexts} directly asked users (crowd-workers and experts) to specify which properties they wanted for different tasks. but, this study measured perceived helpfulness, rather than actualized effects on performance. 
With XAIsim2real, we aim to provide a low-cost way to identify task-relevant properties, from a large set of candidate properties, by computationally simulating their effects on user performance.

\subsection{Properties of XAI methods}
\label{sec: related-work-properties}
Explanation properties originate from a line of ML work that aims to provide automatic evaluations of explanation quality\citep[e.g.][]{liu2021syntheticBenchmarks, bhatt2021evaluating, nguyen2020quantitative, lukasbridging, atanasova2024diagnostic}.
These works rarely validate on real users-- around 20\% of the time, according to the survey by \citet{nauta2023anecdotalQuantitative}. 
Because the proposed property definitions are not grounded in tasks or evaluated on real user performance, there is a lack of consensus on how to formalize and apply these properties.  Disagreements on how to formalize appear across all major types of properties, including complexity (e.g. \citet{bhatt2021evaluating} vs \citet{nguyen2020quantitative}), fidelity \citep{barr2023disagreementFaithfulness}, and robustness (e.g. \citet{yeh2019fidelity} vs \citet{hsieh2020evaluations}). \citet{chen2022harmonized} survey the trade-offs that exist between different properties.  
Our work uses a sim2real approach to test how these different formalizations affect downstream task-performance.

\subsection{Use of proxy users and simulations in XAI}
\label{sec: related-work-proxy}
Several works propose proxy users to identify promising explanation methods before testing on real users. Like us, \citet{chen2022useCaseSimulations} simulate user studies with proxy users to reduce user study costs. Unlike us, their simulations do not account for properties, and as a result, they do not say \textit{why} certain explanations, such as LIME, performed better.  Furthermore, they choose a neural network as their proxy user. We will later show that a more transparent proxy-- whose assumptions about human learning and memory are explicit-- leads to more transparency about how these assumptions affect task-performance. 

Humans proxies have appeared in many other areas of interpretable machine learning. There is a body of works that learns human proxies from human data (e.g. survey responses) and then uses these human proxies to regularize AI functions so that they are easier for humans to work with  \citep{virgolin2020formulaOfInterpretability, hilgard2021learningByHumansForHuman, lage2018human}. Unlike us, these works do not use the human proxies to evaluate explanations. 

\section{Our Goal: Linking Explanation Properties to Tasks}
\label{sec:goal}
In this section, we motivate our goal of linking explanation properties to tasks with two realistic scenarios involving an AI and a human decision-maker. 
Imagine a setting where an AI makes loan recommendations based on an applicant's financial history \footnote{Loan approval is often a topic in XAI \citep{purificato2023Loan1, ustun2019actionableRecourse}, involving popular datasets such as \citet{sharma2023loan}}. 
We consider two distinct decision-making tasks: 
\begin{enumerate}
    \item \textbf{Loan rejection task.} In this scenario, an applicant's loan request has been denied, and they must decide what changes to their application would result in a future acceptance.
    \item \textbf{Compliance task.} In this scenario, an employee must decide whether the AI based its recommendation on illegal attributes about an applicant (such as gender, race, etc). 
\end{enumerate}
Both of these tasks benefit from explanations, but from different \textit{types} of explanation. 
In the loan rejection task, applicants may prefer to focus on only the most influential factors in the AI's recommendation, rather than seeing every detail. 
However, in the compliance task, the employee may prefer to know every detail, no matter how small, so that they can catch even smallest breaches in regulation. 
They also need these details to be accurate, as it would be harmful for an explanation to say that the AI did not rely on an illegal feature when it in fact did. 

Explanation \textit{properties} formalize the extent to which an explanation satisfies the qualities discussed above. 
The property of \textit{complexity} characterizes the amount of information in the explanation-- i.e. level of detail presented to the user. The property of \textit{fidelity} relates to how accurately the explanation reflects the behavior of the true model-- i.e. when the explanation says the AI did not use a feature, it truly did not use it. 
These properties have trade-offs, and the best trade-off depends on the task. 
For the loan-denial task, if the applicant prefers to see only influential factors, then they prefer low-complexity explanations over faithful ones. For the compliance scenario, if the employee needs every detail, then they prefer faithful explanations to low-complexity ones. 

To discover which trade-off between properties is best for a task, user studies are required. In these studies, we would present users with explanations that each satisfy a different trade-off and measure the quality of the resulting decisions. This kind of study is feasible for 2-3 well-defined properties. Unfortunately, the number of explanation properties we care about is large (a recent survey paper lists over 100~\citep{chen2022harmonized}), and user studies quickly become more expensive the more properties we consider. 
The number of properties increases further when considering different ways to formalize characteristics like ``faithfulness.'' One approach measures how much the AI’s recommendations change when the most influential factors are perturbed \citep{dai2022fairness}. Another checks if the explanation provides enough detail to predict the AI’s behavior; without faithfulness, this prediction wouldn't be possible \citep{balagopalan2022road}.
Finally, if we ran a study that identified the important explanation properties in the compliance task, we would have to run yet \textit{another} to identify important properties in the loan rejection task. In short-- the number of possible tasks and properties are both large, and running these studies can be daunting for practitioners for whom this is not their specialty. 

We can assist developers (and researchers with limited resources for user studies) by helping them focus on only the promising properties-- those most likely to improve the user's task-performance. Much of this elimination can happen through domain knowledge; we can hypothesize what kind of explanation would help a user know what financial behaviors to change or to check for compliance. However, the impact of specific choices, such as which formalization of fidelity to use, are more difficult to imagine. This is the role we imagine for XAIsim2real; a tool to narrow the properties included in a user study, when the differences between properties are mathematical and their impact on downstream task-performance is unknown.

\section{Technical Approach: XAIsim2real pipeline}
\label{sec: XAIsim2real}
As noted above in Section~\ref{sec:goal}, different tasks may require explanations with different properties, but identifying the necessary properties for any new task is also a potentially expensive endeavor.  In this section, we describe a way to reduce that cost by using a \emph{sim2real} pipeline in which we use computational proxies of users to first identify promising explanations and then perform user studies on those that are most promising.  
The idea here is not that the computational proxies are somehow perfect representations of real users---indeed, individuals will vary in how they approach tasks---but that computational proxies can still help us rule out explanations that are highly unlikely to be useful and prioritize those that may be. The final step of testing the promising explanations on real users ensures that we are still finding the best explanation for the desired task.

Below, we detail the six components of XAIsim2real, our simulation pipeline for finding those most promising explanations:  
\begin{enumerate}
    \item The \textit{properties} that might be important to the user-decision task 
    \item The \textit{user-decision task} for which the explanations are being used (e.g. checking the loan system for compliance)
    \item The \textit{proxy user model} that will use the explanations to make task-decisions, much like a real user
    \item The \textit{AI function} that needs to be explained 
    \item The \textit{explanation optimization method} to discover explanations that satisfy desirable properties
    \item The \textit{validation study} to check that simulated results transfer to real users 
\end{enumerate}

As we define each component below, we will also demonstrate how we instantiated them for our validation study. 
Though our pipeline is fully general, in our validation study, we assume that explanations take one popular form: feature attributions. Feature attributions are explanations that assign an importance weight for each input feature. For example, if the loan system depended mostly on the user's credit score and not on employment history, then the feature attribution explanation would have a larger number associated with credit score. Throughout, we use $\mathbf{w}$ to refer to the vector of feature attributions that explain the AI function $f$'s decision $y = f(\mathbf{x})$ on the input $\mathbf{x}$.

\begin{figure}[h]
    \centering
    \includegraphics[width=0.7\linewidth]{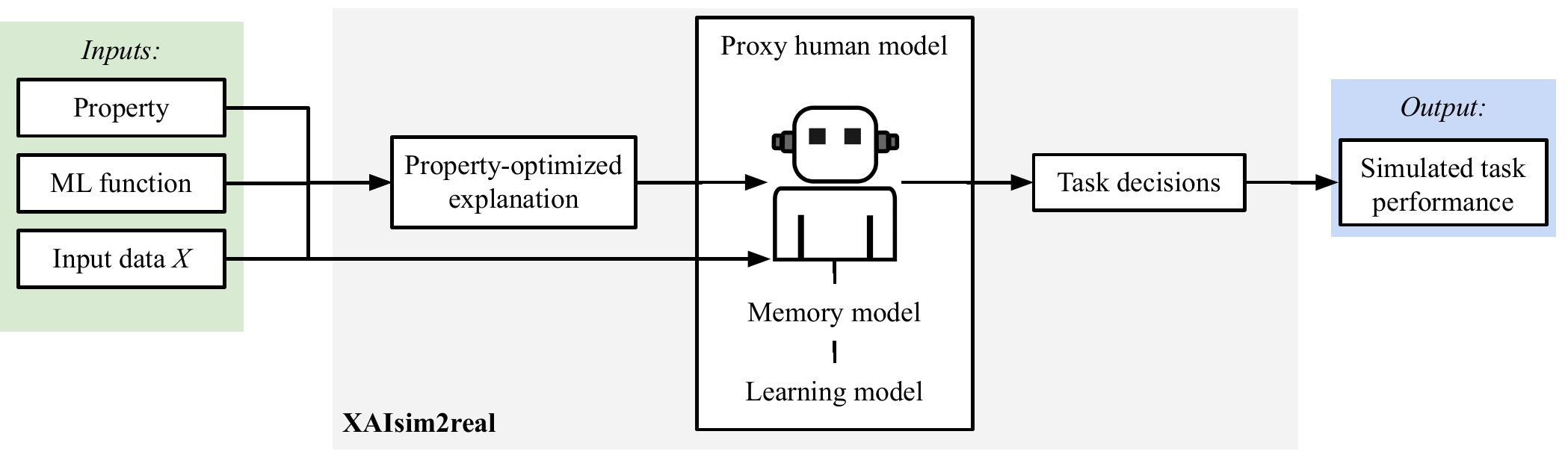}
    \caption{Graphical overview of XAIsim2real, which simulates a user study in three main steps. First, XAIsim2real forms the property-optimized explanations. Second, the user proxy is trained to perform a task on a set of example inputs $\mathbf{x}_h$ and correct decisions $y^*_h$ (the property-optimized explanations are part of the human-inputs). Third, the user proxy is evaluated on a set of test inputs, from which we compute the overall task-performance. }
    \label{fig: XAIsim2real}    
\end{figure}

\subsection{Explanation Properties}
\label{sec: components-properties}
In our pipeline, explanation properties are \textit{functions}, which accept explanations as input and return a number representing the extent to which a characteristic-- such as ``faithfulness''-- is satisfied. 
While all properties are desirable, depending on factors such as the AI function and how the property is formalized, trade-offs between how well an explanation can satisfy each must occur.

\paragraph{Choices for validation study.}
We chose three properties that have already been observed as helpful for different tasks in the literature, but not yet linked to formal property definitions: (1) robustness \citep{jesus2021can}, (2) faithfulness \citep{lertvittayakumjorn2019humanGroundedText, kulesza2013too, nguyen2018comparing}, and (3) complexity \citep{poursabzi2021manipulating, kulesza2013too}. 
For each property below, we chose formalizations from \citet{chen2022harmonized} that work with feature attributions. 

\textbf{Complexity} refers to how much information is contained in an explanation. The human desire for parsimony pushes for simpler and shorter explanations-- thus ones that are less complex. 
Our formalization considers the \textit{sparsity} of the feature attribution, where sparser explanations are less complex. It quantifies sparsity as the number of non-zero weights: 
\begin{equation}
    \label{eq: sparsity}
    \text{sparsity\_loss}(W) = 
    \sum_{d}^{D} \mathbb{I}(\mathbf{w}_d \neq 0),
\end{equation}
where, $D$ is the number of features, $\mathbf{w}_d$ refers to the attribution of the $d$-th feature.

\textbf{Faithfulness} ensures explanations are not misleading by evaluating how well they represent the AI's behavior. 
In making explanations simple (appropriately complex), we make them lose detail and they may become unfaithful to what really happened with the AI (lose faithfulness). 
Our formalization, from \citet{yeh2019fidelity}, measures how well the explanation can recreate the AI's output; explanations that are less faithful will do a worse job at this. 

Our method for using the explanation to recreate the AI output will differ depending on whether the AI is a regression or classification function. 
For AI \textit{regression}, we recreate outputs by weighing each input feature by its corresponding feature attribution: $\mathbf{x}^\top\mathbf{w}$. To measure how faithful the explanation is, we measure the difference between the actual AI output and the recreation. We use the mean-squared error to take this difference: 
\begin{equation}
    \label{eq: fidelity}
    \text{fidelity\_regression\_loss}(\mathbf{x}, \mathbf{w}, p) =  \mathbf{E}_{p(\mathbf{x'} | \mathbf{x})}\left[ (f(\mathbf{x}') - (\mathbf{x}'^\top \mathbf{w}))^2 \right ],
\end{equation}
where $\mathbf{E}$ is the expectation, and $p$ is a distribution over points centered on $\mathbf{x}$.

For AI classification, we still weigh each input feature by its attribution, but we threshold the final value to form a prediction. For binary classification,  we write this as $\ind{\mathbf{x}^\top \mathbf{w} > 0}$, where $\ind{\cdot}$ is an indicator function that evaluates to ``1'' when this weighted sum is greater than zero. We measure the difference between our recreated output and the true AI output by ``counting'' the number of times the two disagree: 

\begin{equation}
    \label{eq: fidelity}
    \text{fidelity\_classification\_loss}(\mathbf{x}, \mathbf{w}, p) =  \mathbf{E}_{p(\mathbf{x'} | \mathbf{x})}\left[ \ind{f(\mathbf{x}') \ne  \ind{\mathbf{x}'^\top \mathbf{w}_n > 0}} \right ].
\end{equation}

\textbf{Robustness} refers to how little an explanation changes with small changes to the input. Explanations that vary too much across inputs are challenging to recall and contextualize, which can cause confusion. This is especially true for similar inputs; they should have consistent explanations to avoid contradictions.

In our formalization from \citet{alvarez2018towards}, if two points, $\mathbf{x}$ and $\mathbf{x}'$,  are within $r$ distance of each other, then they are ``similar,'' and their explanations should not differ significantly:
\begin{equation}
\label{eq: stability}
    \text{robustness\_loss}(\mathbf{x}, \mathbf{w}, r) =     
    \max_{\| \mathbf{x} - \mathbf{x}' \| \le r} \frac{\| \mathbf{w} - \mathbf{w}' \|}{\| \mathbf{x} - \mathbf{x}' \|} .
\end{equation}

Here, $\mathbf{w}$ and $\mathbf{w}'$ are attributions for $\mathbf{x}$ and $\mathbf{x}'$, respectively. The term $\frac{\| \mathbf{w} - \mathbf{w}' \|}{\| \mathbf{x} - \mathbf{x}' \|}$ measures how much the explanations differ relative to how close the inputs are. The maximum value of this ratio over all pairs of points within distance $r$ gives the robustness loss. A higher loss indicates that even small changes in the input can lead to large changes in the explanation, which is undesirable. 

\subsection{User Decision Task}
\label{sec: components-task}
Tasks are the downstream decisions that users make following an explanation of the AI system. 
In our pipeline, tasks describe what information the human uses to make a decision-- which we call the \textit{human inputs} $\mathbf{x}_h$-- and the form of the decision itself, $y_h$. In user studies, we also know the \textit{correct} decision that the human should make, which we refer to as $y_h^*$. Later, we will use pairs of human inputs and correct task decisions, $(\mathbf{x}_h, y_h^*)$, to train proxy humans to perform our tasks. 


\emph{Choices for validation study.}
Like with the explanation properties, we chose tasks that have been studied, but not yet formally linked to properties, in the literature: forward simulation \citep{chen2022useCaseSimulations, poursabzi2021manipulating, lertvittayakumjorn2019humanGroundedText, hase2020evaluatingExplainable}, counterfactual simulation \citep{chen2022useCaseSimulations, hase2020evaluatingExplainable, jeyakumar2020can}, and forbidden features \citep{cornacchia2023auditing, alikhademi2021canUnfairness}. 

In the \textbf{forward simulation} task, the user must predict what the AI function will output for a given input; this task is often used to evaluate a user's understanding of the AI function. Users consider three key pieces information to make task decisions: the AI input $\mathbf{x}$, the explanation $\mathbf{w}$, and how the two interact. Since our explanations are feature attributions, they ``interact'' with the inputs by indicating each feature's influence on the AI's decision. Mathematically, we represent this interaction by multiplying each input feature by its corresponding attribution and adding them together: $\sum_{d = 1}^D \mathbf{w} \cdot \mathbf{x} = \mathbf{w}^\top \mathbf{x}$. The final \textit{human input} is the concatenation of all these pieces: $\mathbf{x}_h = [\mathbf{x}, \mathbf{w}, \mathbf{x}^\top\mathbf{w}]$. 
The \textit{human's decision} takes the same form as the AI's outputs. 
If the AI outputs binary loan decisions $y \in \{0, 1\}$, then the human's decision is also binary, and represents their guess of the loan decision, $y_h \in \{0, 1\}$. The \textit{correct} decision for this task is the actual AI output: $$y_h^* \equiv f(\mathbf{x}).$$

In the \textbf{counterfactual simulation} task, users explore “what if” scenarios where they consider how the AI's output would change if its input were to change. The loan rejection scenario is an example of a counterfactual simulation task; the applicant is curious how changes to their application might affect the AI's loan recommendation. 
The \emph{human's inputs} $\mathbf{x}_h = [\mathbf{x}, \Delta, \mathbf{w}, y]$ include the original AI input $\mathbf{x}$, the ``what if'' changes to the input $\Delta$, the explanation $\mathbf{w}$, and the model's prediction $y$. The $\Delta$ component is a vector whose elements represent how much the corresponding input feature in $\mathbf{x}$ will change. For example, if the first feature represents credit score, and the applicant must decide whether increasing their credit score by ten points would change the AI's recommendation, then $\Delta_1 = 10$, where the subscript refers to the first feature. 
Following the formulation from \citet{jeyakumar2020can} and \citet{chen2022useCaseSimulations}, we restrict this task so that only one feature changes at a time, and by the same amount; the applicant can put the same amount of effort into changing only one aspect of their application. So, $\Delta$ is a scaled one-hot vector, where the position of the non-zero element is random.  
The \emph{human's decision} is one of two possible options-- whether they think the model's prediction for the updated inputs will decrease ($y_h = 0$) or increase/remain the same ($y_h = 1$). The \textit{correct} decision for this task is the direction in which the AI output truly changes: $$y^*_h = \ind{f(\mathbf{x} + \Delta) \ge f(\mathbf{x})}.$$

In the \textbf{forbidden features} task, the user decides whether the AI relied on a forbidden feature, like in the compliance scenario. 
The \emph{human's inputs} $\mathbf{x}_h = [\mathbf{x}, \mathbf{w}, y] $ are the AI input $\mathbf{x}$, the explanation $\mathbf{w}$, and the function's prediction $y$. The \emph{human's decision} is whether the function's output was ($y_h = 1$) or was not ($y_h = 0$) influenced by a forbidden feature, which we fix to be a single feature that is the same for all inputs. The \textit{correct} decision for this task is whether the feature truly affects the model's output: $$y^*_h = \ind{f(\mathbf{x}) = f(\mathbf{x} \text{ without } d)},$$ assuming $d$ is the forbidden feature. 

\subsection{Proxy User Model}
\label{sec: components-proxy}
In user studies, we ask real users to make task-decisions using the explanations. In XAIsim2real, a \textit{proxy} user, $\tilde h$, takes the place of a real user. The goal of the proxy \textit{is not} to perfectly capture human decision-making, but rather, to be a sufficient approximation for identifying which explanations are the most promising. Formally, the proxy user is a function that maps from human inputs $\mathbf{x}_h$ to decisions $y_h$ (note that these are the inputs and decisions defined by the task above). We implement the proxy as a machine learning model trained on pairs of human inputs and correct task decisions. 

While a black-box model, such as a neural network or large-language model, could be used for the proxy, our pipeline includes two subcomponents that expose the proxy user's implicit assumptions on real user \textit{memory} and \textit{learning}. The \textit{memory} model describes how humans process information. While a proxy implemented as a machine learning model has no difficulty processing hundreds of inputs and explanations, real users can only focus on a few. We formalize the memory model as a data pre-proccessing step; it takes in the original human input and outputs a set of modified features. For example, the user's ability to focus on only a few features is a pre-processing step that masks all of the features the user ignores with a $0$.
The \textit{learning} model describes how humans learn to perform a task from experience. While machine learning proxies are trained to perform tasks by optimizing a loss function on data, this process may look different for humans. We formalize the learning model as the optimization algorithm used to train the proxy.

\paragraph{Choices for validation study}
We use a decision tree as our proxy model, because simple, logic-based models are considered interpretable to humans (and therefore thought to mimic their decision-making) \citep{lage2018human}.
We considered other options (a multi-layer perception, K-nearest neighbors) in \cref{appendix: human-model-options}, but we found that our simulation results were not sensitive to the choice of proxy.
Our learning model limits the decision tree depth to two, because we assume people would have trouble grasping deeper logic in one study session. Our memory model rounds human inputs $x_h$ (ranging from 0 to 1) to one significant figure to represent how users might truncate long numbers.

Finally, we consider two versions of the proxy, each with a distinct memory model. The first version, $\tilde h_\text{limited}$, assumes a limited cognitive budget. This limitation is relevant to the \textit{forward simulation} task, where a limited cognitive budget hinders the computation of the inner-product, $\mathbf{w}^\top\mathbf{x}$, that is part of the human's input. In $\tilde h_\text{limited}$, the proxy uses only the two largest feature attribution weights to compute a \textit{partial} inner-product. So, \textit{pre-processed} version of the human input for this task becomes: $\mathbf{x}_h = [\mathbf{x}, \mathbf{w}, \sum\limits_{d = 1}^2 \mathbf{w}^\text{top}_d \cdot \mathbf{x}^\text{top}_d]$, where ``top'' refers to the features with the top-two largest attributions. This represents users who might be overwhelmed when there is a large number input features and choose to focus on the most important ones. 
In the second version, $\tilde h_\text{unlimited}$, the proxy computes the entire inner-product, and $\mathbf{x}_h$ is as defined in \cref{sec: components-task}.

\subsection{AI Functions}
\label{sec: components-functions}
Explanations describe the behavior of an AI's underlying function, $f$, which takes in inputs $\mathbf{x}$ (the loan application) and produces an output $y = f(\mathbf{x})$ (the loan recommendation).
In practice, the AI would already be trained and available for us to query---for example, the bank already has a loan recommendation system whose decisions we must explain. 

\paragraph{Choices for validation study.}  
Instead of learning the AI function $f$ from data, we define it directly; for example, we could define $f$ as a line with fixed weights.  This approach lets us control for variations that would occur from the data and the training process. 

Our choice of $f$ determines whether there are trade-offs in how well it can be explained. For example, suppose we choose an $f$ that is very simple: even though it takes in $100$ inputs, the recommendation only depends on whether the first input is above a certain threshold.  In this case, there is no trade-off: we can faithfully and succinctly explain the entire function just based on the first input. There is no difference between the most sparse, the most faithful, and the most robust explanations---they are all the same.


On the other hand, if $f$ is complex, then there may be trade-offs.  For example, suppose the function is a table that outputs different values for every combination of values for all $100$ features.  Then a \emph{faithful} explanation must include all $100$ inputs because the output truly does depend on all of them.  However, this faithful explanation will not be sparse.  If we need the explanation to be shorter---suppose that the user does not have the capacity to analyze a $100$ weights---then we must exclude some features that impacted the output.  Even if we choose only the most important features to present, we are still showing a simplified (less faithful) explanation.

\begin{figure}[h]
    \centering
    \includegraphics[width=0.6\linewidth]{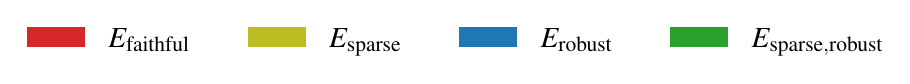}
    
    \includegraphics[width=0.3\linewidth]{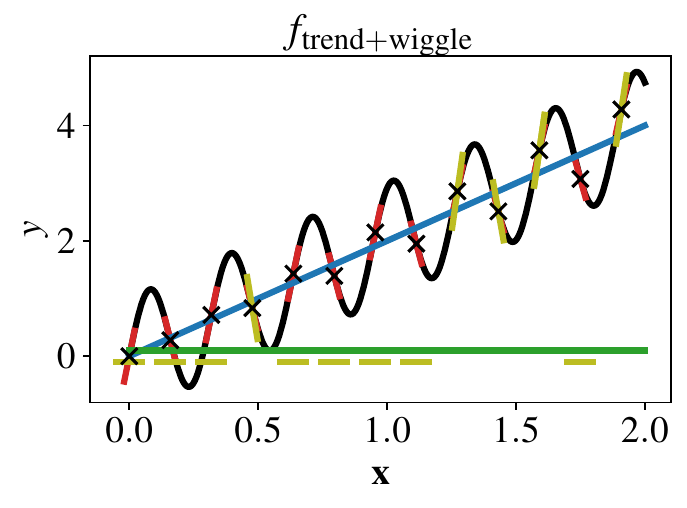}
    \caption{Demonstrative, one-dimensional example of $f_\text{trend+wiggle}$, and how explanations with different properties would describe it differently. This function (in black) has a clear long-term linear trend with short-term oscillations on top. Explanations are represented as lines at input points of interest (`X'). Faithful explanations (red) will capture the short-term oscillations. Sparse explanations (yellow) will also do this, but will also return `0' at points, to maintain sparsity. Robust explanations (blue) will capture the long-term trend but not the oscillations. Sparse+robust explanations (green) are zeroed-out for this dimension, but can be non-zero for other dimensions.}
    \label{fig: f-trend-wiggle}
\end{figure}

During our initial exploration, we learned that choosing AI functions with trade-offs was important to creating situations with signal (different properties cause differences in task-performance). 
So, we chose AI functions that we \textit{knew} would induce trade-offs. 
While these functions appear abstract, we believe each of them categorizes a different ``type'' of realistic AI behavior, as we describe below:
\begin{itemize}
    \item The first function, $f_\text{sparse}$, is designed so that a sparse explanation that selects only one feature to present to the user is still faithful, but the feature to be selected varies with the input (not robust). For example, the loan system only considers the applicant's credit score or savings account in making a recommendation, but whether the explanation places more emphasis on credit score or savings depends on the applicant. 
    \item The second function, $f_\text{trend+wiggle}$ induces a trade-off between faithfulness and robustness. 
    This trade-off is best understood by looking at the shape of the function in \cref{fig: f-trend-wiggle}; an explanation that is as faithful as possible will be non-robust, because it must capture all of the local ``wiggles.''
    Likewise, an explanation that is robust will not be as faithful, because it ignores the local variation to capture the long-term trend.
    This kind of function represents an AI loan system whose recommendations are highly-specific to each applicant, but some general trends describe successful applicants (e.g. a high credit score). 
    \item The third function, $f_\text{wiggle}$, is like $f_\text{trend + wiggle}$, without the clear long-term trend. As a result, the trade-off between complexity, robustness, and faithfulness is not as clear. This function represents complicated loan systems where the reasoning behind the decision is highly-specific to each applicant.
\end{itemize}
Definitions of each function are in \cref{appendix: ai-functions}.


\subsection{Method for Optimizing Explanations to Properties}
\label{sec: components-property-optimized-explanations}
Properties are attributes of explanations. In our paradigm, humans interface with explanations, rather than with the properties directly, to make decisions. 
XAIsim2real optimizes explanations to satisfy a given property for the user to interface with. For example, if we want to find a faithful explanation, we would search for explanations that minimize the faithfulness loss in \cref{eq: fidelity}. In general, since properties are functions of explanations, they can be used as optimization objectives. 

While it is possible to find explanations that optimize for other properties, like sparsity, these explanations are not useful if they are not faithful to some degree. For example, an explanation that is \textit{perfectly sparse} and \textit{unfaithful} would assign an attribution of $0$ to all input features. This kind of explanation contains no information that can be used to make a good decision regarding the underlying AI function. 
So, we are interested finding explanations that satisfy one property while also being \textit{as faithful as possible}, which we pose as the following optimization problem: 
\begin{align}
    \label{eq: optimization_problem}
    \min_{\mathbf{w}} \text{faithfulness\_loss}(\mathbf{x}, \mathbf{w}, p)\\
    \textbf{subject to } \text{prop\_loss}(\mathbf{x}, \mathbf{w}, \ldots) = 0.
\end{align}
Here, the primary concern is satisfying the desired property (captured by ``prop\_loss'').  Once this property is met, we then focus on making the explanation as faithful as possible (captured by ``faithfulness\_loss''). 
The generic reference to ``prop\_loss'' can be replaced by any property, such as robustness \cref{eq: stability}. 

\paragraph{Choices for validation study.} 
In our study, the property-optimized explanations are the independent variable. 
Since we ensure our explanations are always \textit{as faithful as possible}, only sparsity and robustness remain as free variables from our list of properties in  \cref{sec: components-properties}. 
As a result, we consider four types of property-optimized explanations: explanations that are neither sparse nor robust ($E_\text{faithful}$), sparse but not robust ($E_\text{sparse}$), robust but not sparse ($E_\text{robust}$), and both sparse and robust ($E_\text{sparse+robust}$). In \cref{tab: example-explanations}, we provide examples of the optimized feature attributions for function $f_\text{sparse}$. 

\newcommand{\fm}{\phantom{-}}
\begin{table}[h]
    \centering
    \small
    \begin{tabular}{r| c | c}
    \toprule
        & 
        $\mathbf{x}_1$ & 
        $\mathbf{x}_2$\\
        \midrule
        Input  & 
            $\begin{pmatrix} \fm0.70 &  \fm0.50 & \fm0.80 & \fm0.10 \end{pmatrix}$ & 
            $\begin{pmatrix}  \fm0.50 & \fm0.10 &  \fm0.40 & \fm1.00 \end{pmatrix}$ \\ \midrule
        $E_\text{faithful}$ & 
            $\begin{pmatrix} -1.00 & -0.96 &  \fm0.60 & \fm0.50 \end{pmatrix}$ & 
            $\begin{pmatrix} \fm0.00 & -1.00 &  \fm0.60 & \fm0.50 \end{pmatrix}$ \\
        $E_\text{robust}$ & 
            $\begin{pmatrix} -0.85 & -0.82 &  \fm0.12 & \fm0.74 \end{pmatrix}$ & 
            $\begin{pmatrix} -0.85 & -0.82 &  \fm0.12 & \fm0.74 \end{pmatrix}$ \\
        $E_\text{sparse}$ & 
            $\begin{pmatrix} -4.00 &  \fm0.00 &  \fm0.00 & \fm2.00 \end{pmatrix}$ & 
            $\begin{pmatrix}  \fm0.00 & -1.00 &  \fm0.00 & \fm0.50 \end{pmatrix}$ \\
        $E_\text{sparse+rob.}$ & 
            $\begin{pmatrix} -1.00 &  \fm0.00 &  \fm0.00 & \fm0.57 \end{pmatrix}$ & 
            $\begin{pmatrix} -1.00 &  \fm0.00 &  \fm0.00 & \fm0.57 \end{pmatrix}$ \\
    \bottomrule
    \end{tabular}
    \caption{Example of what each property optimized explanation of $f_\text{sparse}$ looks like for two different inputs, one input per column. The robust explanations ($E_\text{robust}$ and $E_\text{sparse+robust}$) do not change across the two different inputs, and the sparse explanations ($E_\text{sparse}$ and $E_\text{sparse+robust}$) have more zero terms than the non-sparse explanations. }
    \label{tab: example-explanations}
\end{table}

To optimize for explanations that are sparse \textit{and} robust, the optimization problem posed in \cref{eq: optimization_problem} would be subject to two constraints, one for sparsity (\cref{eq: sparsity}) and the other for robustness (\cref{eq: stability}). To optimize for explanations that are neither, there would be \textit{no constraint}, and the problem would be reduced to one that maximizes faithfulness. 
Note that we have to conduct these optimizations separately for each AI function. 
When we can, we use our knowledge of the AI function to optimize. For example, to optimize for faithfulness when the underlying function is linear around an input, we return, as our explanation, the ground-truth weights of the linear function. Such explanations will be optimally faithful under any formalization. 
On properties and functions where we cannot use the AI function, we randomly sample feature attributions and select ones that result in the best loss. The breakdown of which explanation-function combinations we sample is in \cref{appendix: explanation-optimization}.
We leave the fully general task of optimizing explanations for arbitrary (possibly unknown) functions as future work.

\subsection{Validation Study}
\label{sec: components-validation-study}
Once XAIsim2real is used to run synthetic user studies, the most promising explanations should be validated on real users. To validate, we need to design a user study that mirrors the synthetic one. Like with our proxy users, real users should learn to use the explanations to perform a task in a training phase, and then be evaluated on how well they make task-decisions in a test phase.

However, pipeline components such as ``AI functions'' are too abstract and not interpretable to real users. Each component should be presented in a user-friendly way (e.g. ``the AI function is a loan-recommendation system''), while minimizing differences from the synthetic study-- if the synthetic study used $f_\text{sparse}$, then the loan-recommendation system's decisions should be provided by $f_\text{sparse}$. 
In the next two sections, we demonstrate our process for running the synthetic \cref{sec: simulated-user-experiments} and real \cref{sec: real-user-experiments} studies. 

\section{Synthetic user experiments: hypotheses generation}
\label{sec: simulated-user-experiments}
In this section, we use the XAIsim2real pipeline to generate hypotheses about which properties our users will prefer, which we will then test in \cref{sec: real-user-experiments}.
Our \textit{primary independent variable} is the property-optimized explanation. 
The \textit{dependent variable} is the proxy user's task-performance. 
The remaining pipeline components-- the AI functions, tasks, and proxy user models-- are additional factors that, when combined, form the \textit{experiment condition}. 

We suggest the following three steps for generating hypotheses using XAIsim2real: 
\begin{enumerate}
    \item \textbf{Exploration}. Since we do not know which combinations of pipeline components will result in situations where we will see separation of task-performance by property, we run a synthetic study for all possible experimental conditions. 
    \item \textbf{Selection}. We select the experimental conditions that result in performance differences among the explanation types. 
    \item \textbf{Scaling back}. We re-run the selected conditions on a small-scale synthetic study (one with a smaller number of proxy users and test questions) that is feasible to conduct with real users. We do this because we expect effect size to reduce with a smaller study, and we want to see if significant differences can still be measured. Some further guidance on how to choose train and test instances for the small-scale study: 
    \begin{itemize}
        \item \textit{Training instances.} To ensure users can learn necessary strategies from a small number of examples, training should include pedagogical instances that inform the user about a concept important to the task. For example, in the forbidden features task, we may include one example where the attribution at the forbidden feature is $0$ (so the user should recognize the correct decision is ``forbidden feature \textit{was not} used) and another example where everything else is the same, but the attribution is non-zero (so the user should recognize the correct decision is ``forbidden feature \textit{was} used). 
        \item \textit{Test instances.} We select the instances most likely to show the effect (i.e. that the hypothesized ``best explanation type'' holds). These are instances from the exploration phase where the proxy user made the correct decision under the best explanation type but made incorrect decisions when given others. To protect against situations where the proxy user is a poor stand-in for real users, we also include instances where the proxy users made the same decision, regardless of explanation type and instances where the proxy users made \textit{incorrect} decisions with the best explanation type. 
    \end{itemize}
\end{enumerate}

Below, we detail the results of applying these steps to our validation study.

\subsection{Setup for synthetic experiments}
Our property-optimized explanations-- the primary independent variable-- takes on the four types from \cref{sec: components-property-optimized-explanations}: $E_\text{faithful}$, $E_\text{robust}$, $E_\text{sparse}$, and $E_\text{sparse+rob}$. These explanation types are controlled for confounding; explanations that are not optimized to be robust have similar levels of baseline robustness, and so on. 

The user's task-performance-- the dependent variable-- is measured as the proxy model's accuracy on the test points. The remaining pipeline components are factors that form the experimental conditions. For us, this is a $3 \times 3 \times 2$ experiment design with $18$ conditions in all: $3$ tasks (forbidden features, counterfactual simulation, forward simulation), $3$ functions ($f_\text{sparse}$. $f_\text{trend+wiggle}$, $f_\text{wiggle}$), and $2$ proxy users ($\tilde h_\text{limited}, \tilde h_\text{unlimited}$).

The pipeline requires specifying a set of training and testing data points. The training points are used to optimize the explanations and to form part of the inputs for training the human proxy. For optimizing the explanations, we sampled $500$ points from near the AI function's decision boundary, because points near the boundary are the most informative of the function's behavior. For training the human proxy, we sampled $10$ of these points. 
We used such a small number, because this was the scale of training data we expected to show a human in a one-sitting during a user study.
Test points are a set of $500$ evenly spaced input points.
For the scaled-back study, we chose $10$ training points and $30$ test points (details in \cref{appendix: train-and-test-selection}).

\subsection{Results}
\label{sec: simulated-user-results}
\paragraph{Exploration: running a synthetic study for all combinations}

\begin{figure}[h]
    \centering
    \begin{subfigure}{1\linewidth}
        \includegraphics[width=0.6\linewidth]{figures/sim-legend.pdf}
    \end{subfigure}
    \begin{subfigure}{0.33\linewidth}
        \includegraphics[width=1\linewidth]{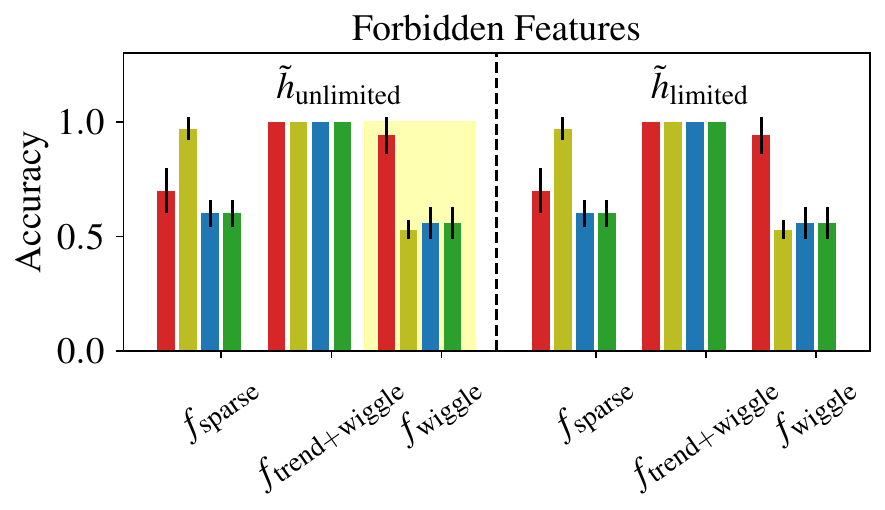}
        \caption{}
        \label{fig: simulation-forbidden-features}
    \end{subfigure}%
    \begin{subfigure}{0.33\linewidth}
        \includegraphics[width=1\linewidth]{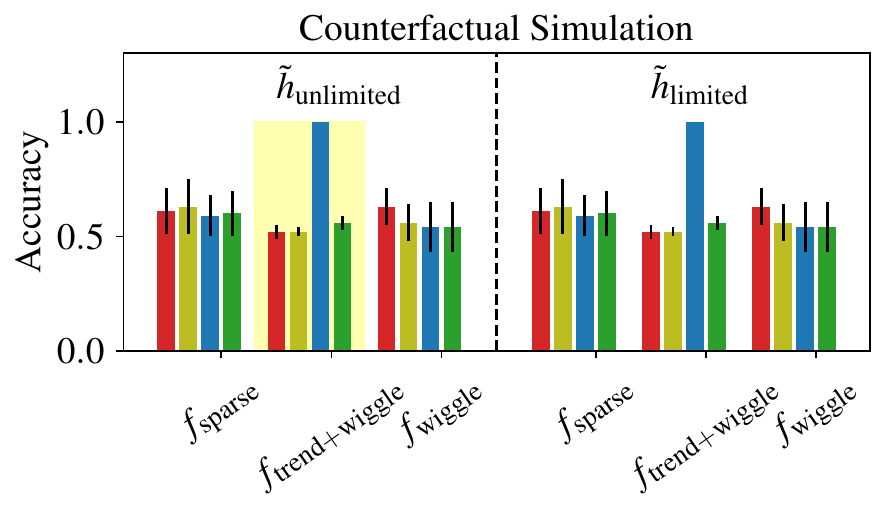}
        \caption{}
        \label{fig: simulation-counterfactual-simulation}
    \end{subfigure}
    \begin{subfigure}{0.33\linewidth}
        \includegraphics[width=1\linewidth]{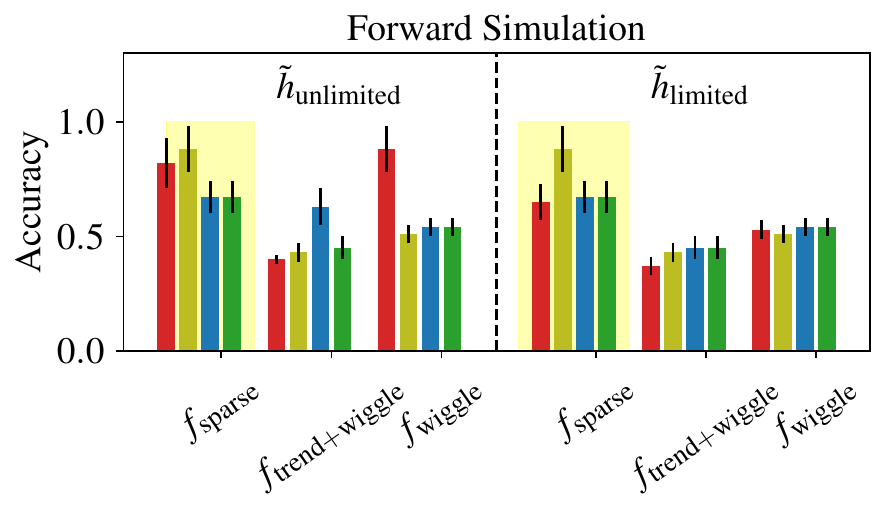}
        \caption{}
        \label{fig: simulation-forward-simulation}
    \end{subfigure}
    \caption{Simulation results. Each subplot is a task and each grouping represents results within a setting. We validate the settings highlighted in yellow with user studies. 
    Error bars are 95\% confidence intervals over the different proxies that result from resampling the $10$ training points. 
    }
    \label{fig: simulation-results-all}
\end{figure}

In \cref{fig: simulation-results-all}, we present the results of our eighteen synthetic study conditions. Each task shows a propensity toward a certain type of explanation: 
\begin{itemize}
    \item \textit{Faithful explanations for forbidden features.} In \cref{fig: simulation-forbidden-features}, faithful explanations were best. This is most apparent on $f_\text{wiggle}$, where the dependency on the forbidden feature is often small but non-zero (an attribution of $0.05$), and faithful explanations could capture this dependency. For $f_\text{sparse}$, sparse explanations are also perfectly faithful, and do well on this task. 
    All explanations perform equally well in $f_\text{trend+wiggle}$ because this function relies on all of the inputs; the forbidden feature is always used, so there is not a decision to make. 
    \item \textit{Robust explanations for counterfactual simulation.} 
    Consider \cref{fig: simulation-counterfactual-simulation}, where robust explanations were best for $f_\text{trend+wiggle}$. In this task, the user had to consider how the AI function would extrapolate to new cases, so explanations that provided a global understanding of the AI's behavior were helpful. While faithful explanations captured the local wiggles in $f_\text{trend+wiggle}$, they failed to explain the main trend to the user. In contrast, robust explanations captured the main trend instead of the local wiggles. 
    \item \textit{Faithful explanations (with complexity considerations) for forward simulation.} Unlike the other tasks, in forward simulation (\cref{fig: simulation-forward-simulation}), modeling the human with a limited cognitive budget affected which property was preferred. This difference is because forward simulation with feature attributions requires several mathematical operations; one has to multiply the inputs by the corresponding attributions, and then add them together to make a decision. Our simulation findings indicate that people with a limited cognitive budget will prefer sparse explanations, which require them to perform fewer operations overall. In contrast, people with no budget limitations will perform equally well with faithful explanations, which require more operations to arrive at the correct answer. 
\end{itemize}

\paragraph{Selection: narrowing in on promising conditions}
We selected the synthetic study conditions with significant effects; these are the groups in \cref{fig: simulation-results-all} with non-overlapping differences in performance among explanation types. 
While all ten of the conditions with effects in \cref{fig: simulation-results-all} had the potential to be interesting, we selected three conditions that highlighted a different property for each task. We also added a fourth condition that demonstrated the effect of the cognitive budget: 
\begin{enumerate}
    \item Forbidden features with $h_\text{unlimited}$ and $f_\text{wiggle}$ to highlight faithful explanations
    \item Counterfactual simulation with $h_\text{unlimited}$ and $f_\text{trend+wiggle}$ to highlight counterfactual explanations
    \item Forward simulation with $h_\text{unlimited}$ and $f_\text{sparse}$ to highlight sparse \textit{and} faithful explanations
    \item Forward simulation with $h_\text{limited}$ and $f_\text{sparse}$ to highlight preference for sparse explanations under cognitive budget constraints
\end{enumerate}

To not duplicate results, we omitted conditions where limitations to the cognitive budget did not result in different effects (e.g. the right and left side of \cref{fig: simulation-forbidden-features} are the same). 
We omitted forbidden features with $f_\text{sparse}$ and $\tilde h_\text{unlimited}$ because we already selected a condition where sparse explanations were preferred in forward simulation. We omitted forward simulation with $f_\text{trend+wiggle}$ and $\tilde h_\text{unlimited}$ because the effect size was small. We do not further investigate non-significant effects from the synthetic study, since they are hypothesized to be null.

\begin{figure}[h]
    \centering
    \begin{subfigure}{0.25\linewidth}
        \includegraphics[width=1\linewidth]{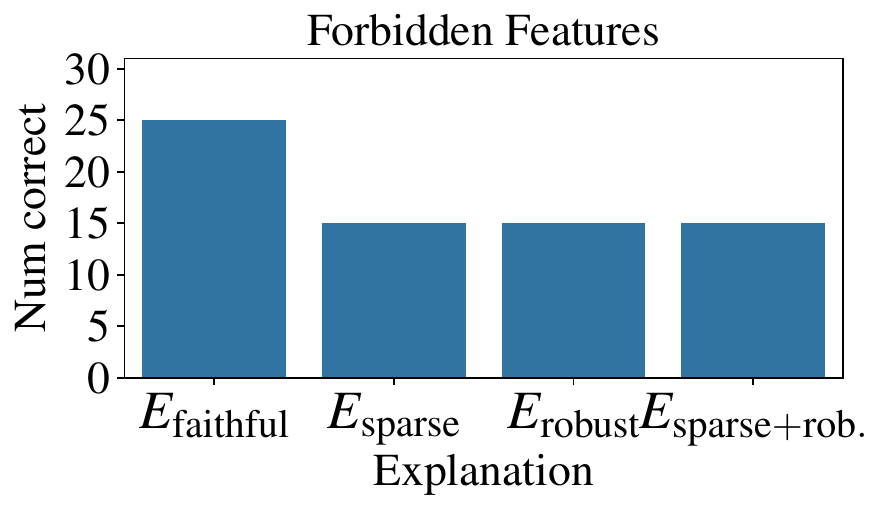}
        \caption{}
        \label{fig: small-scale-forbidden-features}
    \end{subfigure}%
    \begin{subfigure}{0.25\linewidth}
        \includegraphics[width=1\linewidth]{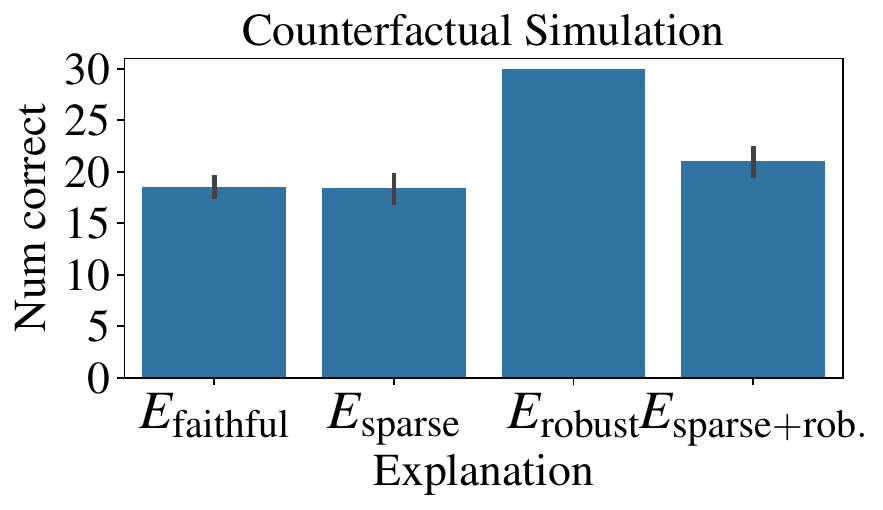}
        \caption{}
        \label{fig: small-scale-counterfactual-simulation}
    \end{subfigure}%
    \begin{subfigure}{0.25\linewidth}
        \includegraphics[width=1\linewidth]{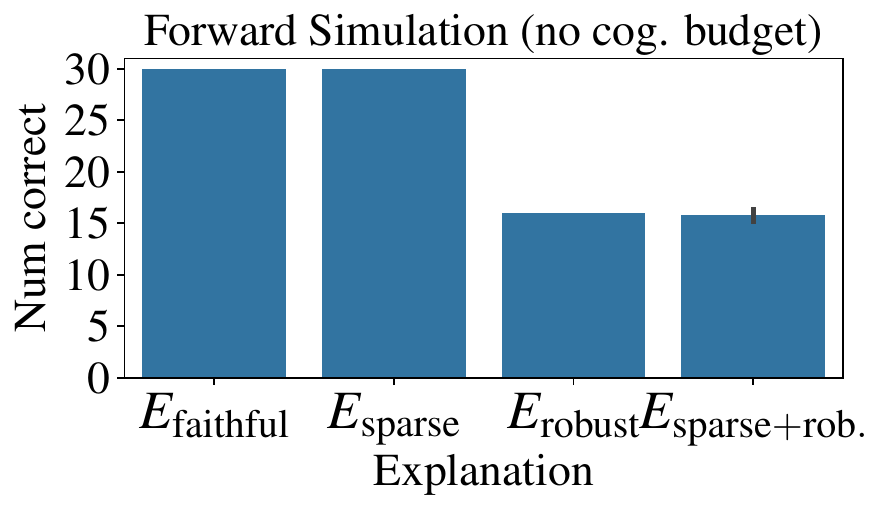}
        \caption{}
        \label{fig: small-scale-forward-simluation}
    \end{subfigure}%
    \begin{subfigure}{0.25\linewidth}
        \includegraphics[width=1\linewidth]{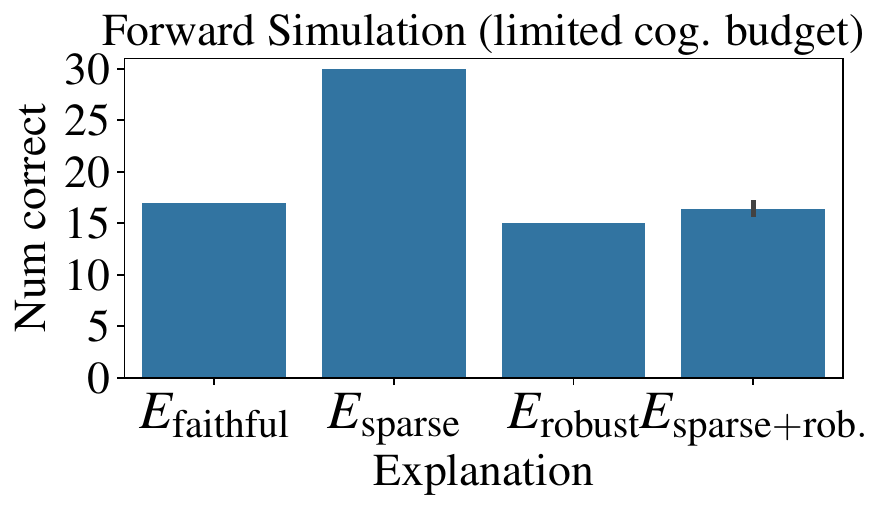}
        \caption{}
        \label{fig: small-scale-forward-simluation-tp}
    \end{subfigure}
    \caption{The proxy users used the same $10$ training points and $30$ test points as the real users. Error bars are 95\% confidence intervals across training the model (randomness due to the initialization of the decision tree optimization).
    }
    \label{fig: simulation-results-selected}
\end{figure}

\paragraph{Scaling back: checking that effects still hold with fewer data.}
In \cref{fig: simulation-results-selected}, we show the result of re-running the four conditions on a smaller set of $10$ training points and $30$ test points (specific to each task). The same relationships from the large-scale study held: $E_\text{faithful}$ was best for forbidden features, $E_\text{robust}$ was best for counterfactual simulation, $E_\text{faithful}$ and $E_\text{sparse}$ were equally good for forward simulation without time pressure, and $E_\text{sparse}$ was best for forward simulation with time pressure. 
The only notable difference was that, on counterfactual simulation, $E_\text{sparse+rob.}$ outperformed $E_\text{sparse}$ and $E_\text{faithful}$. 


\section{Real user experiments: Hypotheses testing}
\label{sec: real-user-experiments}
In the prior section, we used the proxy user to hypothesize about property-task relationships. In this section, we conducted user studies to test these hypotheses. 

While our proxy users are not a perfect representation of real users, we expect them to be sufficiently reliable for ranking the relative performance of explanation properties within a task. That is, we hypothesized that if proxy users indicated that one explanation is better than another, then real users would prefer the same explanation over another, or the difference between explanations would be indistinguishable due to noise. 

We further expected that the best explanation property would vary by task, with the following hypotheses about which would be best:
\begin{itemize}
    \item [\textbf{H1}] Participants will perform best on the \textit{forbidden features} task with \textit{faithful} explanations.
    \item [\textbf{H2}] Participants will perform best on the \textit{counterfactual simulation} task.
    \item [\textbf{H3}] Participants will perform equally well on the \textit{forward simulation} task with \textit{faithful} and \textit{sparse} explanations.
    \item [\textbf{H4}] Participants will perform best on the \textit{forward simulation} task with \textit{sparse} explanations when there is time pressure (i.e. when participants are subject to a limited cognitive budget). 
\end{itemize}

Our four hypotheses are to validate that the ``best'' explanation found on our proxy users is also best for real users. In our synthetic study from \cref{sec: simulated-user-results}, we noted several effects where one explanation type was preferred over another, even though neither was the best within the task. We list the full set of significant effects from the synthetic study in \cref{tab: proxy-vs-real-relationships}. 

\subsection{Experiment apparatus}
\label{sec: real-experiments-apparatus}
Since our goal was to test whether the same relationships from the simulation would hold on real users, we designed the user study to mirror the synthetic on as much as possible. However, aspects of the simulation needed to be presented in a user-friendly manner. For example, rather than presenting the input data $\mathbf{x}$ as a vector of numbers, we attached a semantically meaningful to each feature.

We created three toy scenarios for users to interact with, one for each task.
The toy scenarios were identical to the tasks in \cref{sec: components-task}, except that we presented the information (such as the inputs, AI functions, and user decisions) in a way that was engaging and accessible for real users.
In our toy scenarios, participants provided medical treatment to extraterrestrial aliens, as in \citet{lage2019evaluationAliens} and \citet{swaroop2024timePressure}. We chose this alien scenario, rather than a more realistic one (such as predicting house prices), to mitigate confounding from different levels of task-specific prior knowledge -- i.e. humans have to rely purely on the explanation and our training, rather than prior knowledge of alien physiology. Furthermore, to mitigate confounding from different levels of trust in an ``AI,'' we removed all mentions of an ``AI'' and referred to ``AI explanations'' as information from an ``alien researcher.'' 

\begin{figure}[h]
    \centering
    \includegraphics[width=0.5\linewidth]{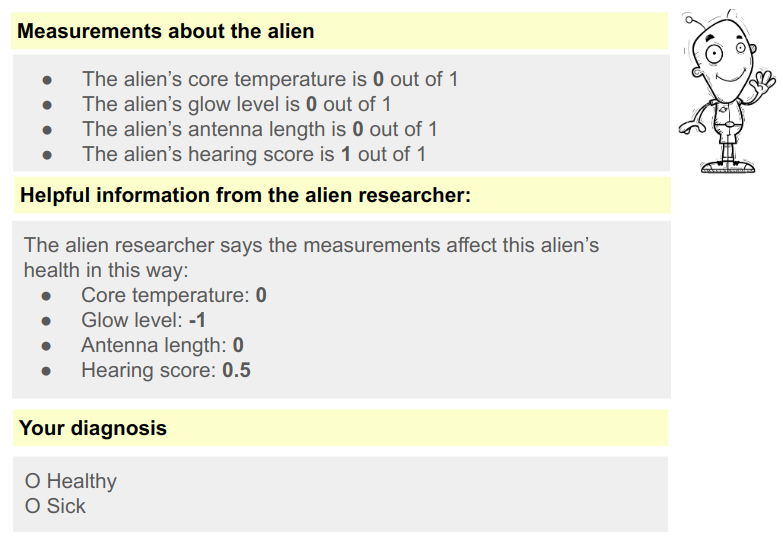}
    \caption{Example UI for forward simulation task, where participants diagnosed alien as healthy or sick. The explanation (covered as helpful information from an alien researcher) is from a sparse explanation. UIs for the remaining tasks are in \cref{appendix: screenshots}.   
    }
    \label{fig: alien-UI}
\end{figure}

For \textbf{forward simulation}, participants diagnosed an alien with imaginary physical traits (e.g. glow). The diagnosis was a binary label: healthy or not healthy. The AI function determined the mapping from physical traits to a diagnosis. Explanations were presented as an ``alien researcher's'' advice on how physical traits affect alien health. 

In \textbf{counterfactual simulation}, participants decided whether an alien patient's medical risk score, which was originally assigned at intake, had gone up or down now that a measurement had changed. The AI function defined the risk score. Explanations were presented as an ``alien researcher's'' advice on how measurements affect risk score. 

In \textbf{forbidden features}, participants decided whether a doctor relied on a forbidden trait to diagnose aliens. The AI function determined whether the doctor truly used the trait for their diagnosis. Explanations were presented as an ``alien researcher's'' opinion on which traits the doctor used. The forbidden trait remained the same across all participants and instances.

Finally, we used time pressure to encourage the limited cognitive budget condition of one of our proxy users. Time pressure has been used in cognitive psychology to create situations in which individuals must make decisions with limited cognitive resources, because it forces them to prioritize certain information while ignoring other details \citep{keinan1999effectStress}. During the test phase of the study, participants saw a global timer for how long they had to complete all questions and a local timer for a ``recommended'' time per task (total time divided by total number of questions); this procedure followed the one used in \citet{swaroop2024timePressure}. 

\subsection{Methods}
\label{sec: real-experiments-methods}
Our study was conducted online on Prolific. We hosted each task as a separate experiment on the platform, so participants were randomly allocated to the tasks (between-subject).  
We recruited $287$ total participants across tasks. Of the participants, $156$ passed comprehension checks. 
We used a between-subjects design, where each participant was randomized to an explanation type (either $E_\text{faithful}$, $E_\text{sparse}$, $E_\text{robust}$, or $E_\text{sparse+rob}$). 
In total, we had a median of $10$ participants per condition (min $7$, max $13$), across four hypotheses and four explanation types per hypothesis. 
Participants who completed the study were $56\%$ percent male, $54\%$ percent belonged to the $18 - 34$ age range, and were located in the United States. 
Participants were paid $\$12$ per hour (median time $18$ minutes) with a \$1 bonus for the best performance within a condition and a \$1 bonus for the best-written responses. 
Our study was approved by the Internal Review Board at [university].

\subsection{Procedure}
\label{sec: real-experiments-procedure}
Before the study, participants accepted a consent form, and after the study, they filled out an exit survey that included a manipulation check and likert questions about whether they enjoyed puzzles and math. 

The main study had four phases: (1) instructions, (2) comprehension checks, (3) task training, and (4) task testing. 
In the \textbf{instruction}, participants were introduced to the alien scenario and the UI. Participants were told what the explanation means (``an alien researcher will give you their best understanding of how each alien's measurements affect its overall health''), along with how to interpret them (``higher [feature] is \textit{worse} for alien health, because this number is negative,'' when describing a feature with a negative attribution). They were not instructed on how to use explanations to make decisions.
The \textbf{comprehension check} questions assessed the participant's basic understanding of the inputs and explanations. For example, for forward simulation, we asked, ``According to the alien researcher, which measurement would have the biggest effect on the alien’s health?,'' to check that the participant knew that a higher absolute value meant a higher feature attribution. 
In the \textbf{training phase}, participants were instructed to form a decision-making strategy from $10$ training instances, which included correct decisions. They were asked to write down their strategy before moving on to the final phase. 
Finally, in the \textbf{test phase}, participants applied their strategy to $30$ test instances (order randomized). 

\subsection{Analysis}
\label{sec: real-experiments-analysis}
Our performance metric is the accuracy on $30$ questions.  
We also considered the participant's written decision-making strategies from the training phase as part of our qualitative analysis. 

We used a one-sided Brunner-Munzel test to perform pairwise comparisons of explanation properties, with a Bonferroni correction to account for multiple comparisons. This test is appropriate because our response variable is discrete, ordinal, and we expected the variances between groups to differ, depending on the helpfulness of the explanation.

For analyzing participant's written strategies, a single researcher conducted an initial open coding phase by reviewing all participant responses across tasks to identify a core set of codes. Following this, the researcher employed a deductive coding approach, systematically applying the predefined codes to each response. A single response could belong to more than one code. 
To facilitate interpretation, we grouped the codes by strategy type: strategies with \textit{reasonable} use of explanation, strategies with \textit{misuse} of explanation, strategies that \textit{ignore} the explanation, and those with \textit{no strategy}. A ``reasonable'' use of explanation differed by task (full groupings and codes in \cref{appendix: codes}).

\subsection{Results}
\label{sec: real-experiments-results}
\begin{figure}[h]
    \centering
    \includegraphics[width=0.8\linewidth]{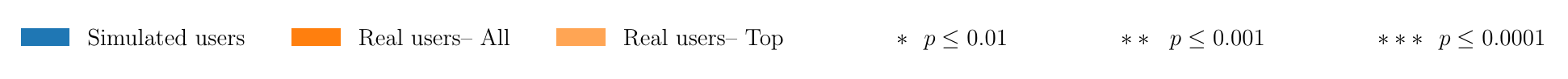}
    All users
    \begin{subfigure}{0.25\linewidth}
        \includegraphics[width=1\linewidth]{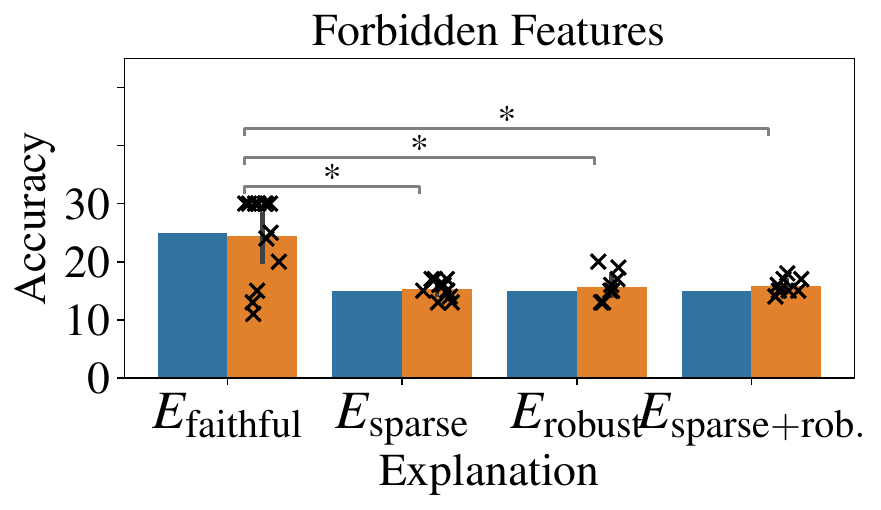}
        \caption{}
        \label{fig: human-forbidden-features-all}
    \end{subfigure}%
    \begin{subfigure}{0.25\linewidth}
        \includegraphics[width=1\linewidth]{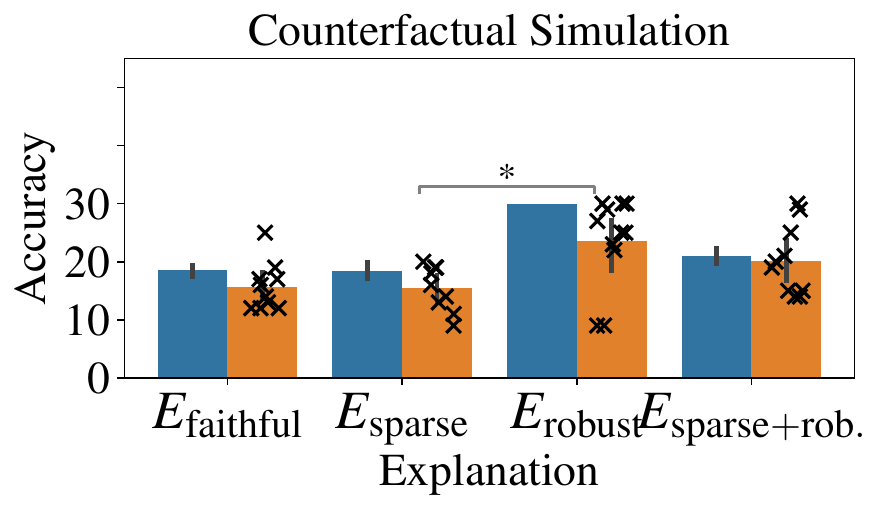}
        \caption{}
        \label{fig: human-counterfactual-sim-all}
    \end{subfigure}%
    \begin{subfigure}{0.25\linewidth}
        \includegraphics[width=1\linewidth]{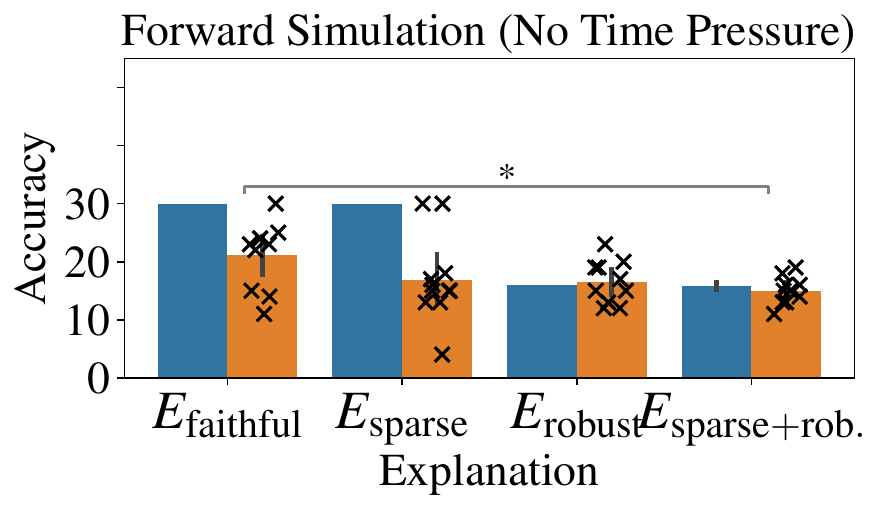}
        \caption{}
        \label{fig: human-forward-no-pressure-all}
    \end{subfigure}%
    \begin{subfigure}{0.25\linewidth}
        \includegraphics[width=1\linewidth]{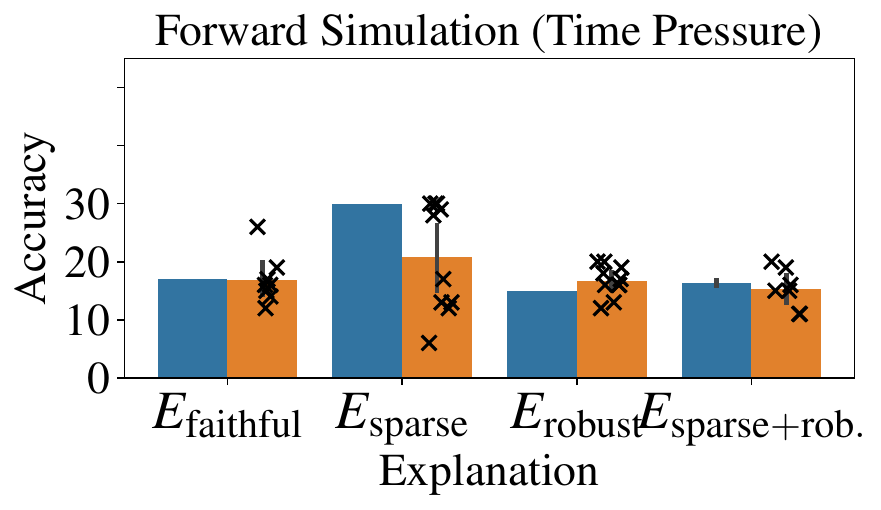}
        \caption{}
        \label{fig: human-forward-pressure-all}
    \end{subfigure}
    Top 50\% of users
    \begin{subfigure}{0.25\linewidth}
        \includegraphics[width=1\linewidth]{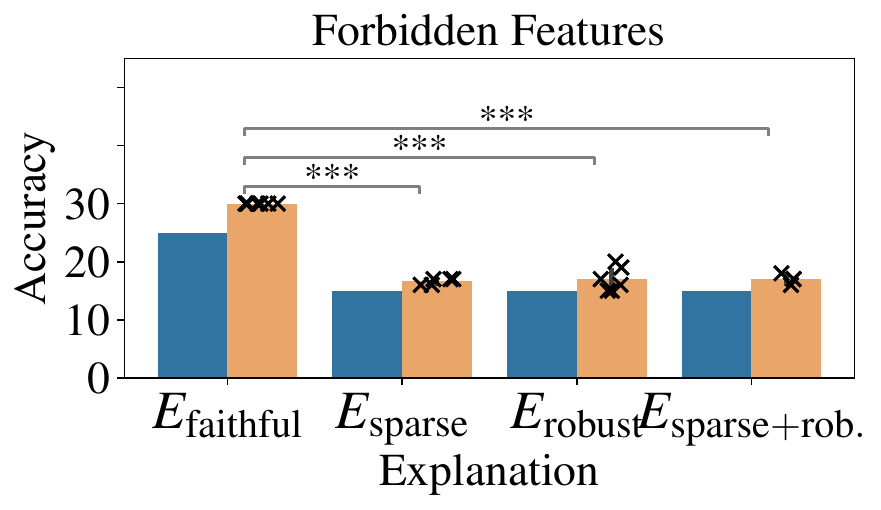}
        \caption{}
        \label{fig: human-forbidden-features-top}
    \end{subfigure}%
    \begin{subfigure}{0.25\linewidth}
        \includegraphics[width=1\linewidth]{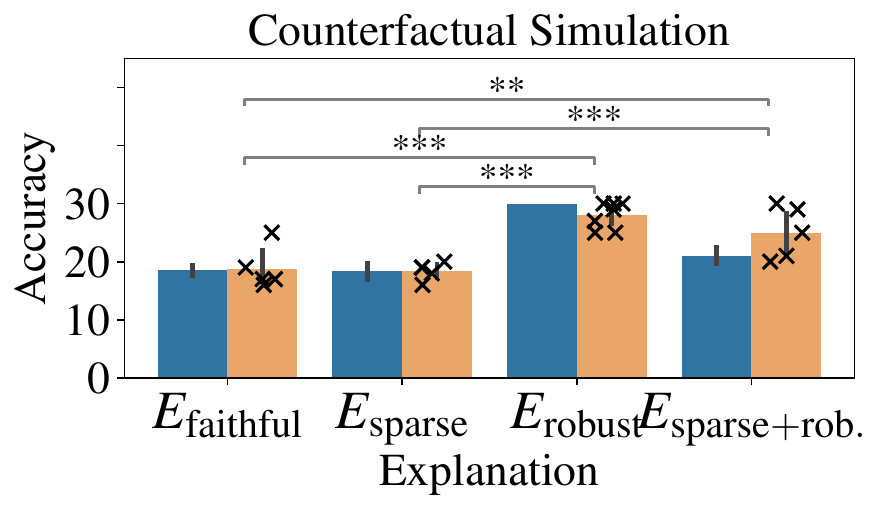}
        \caption{}
        \label{fig: human-counterfactual-sim-top}
    \end{subfigure}%
    \begin{subfigure}{0.25\linewidth}
        \includegraphics[width=1\linewidth]{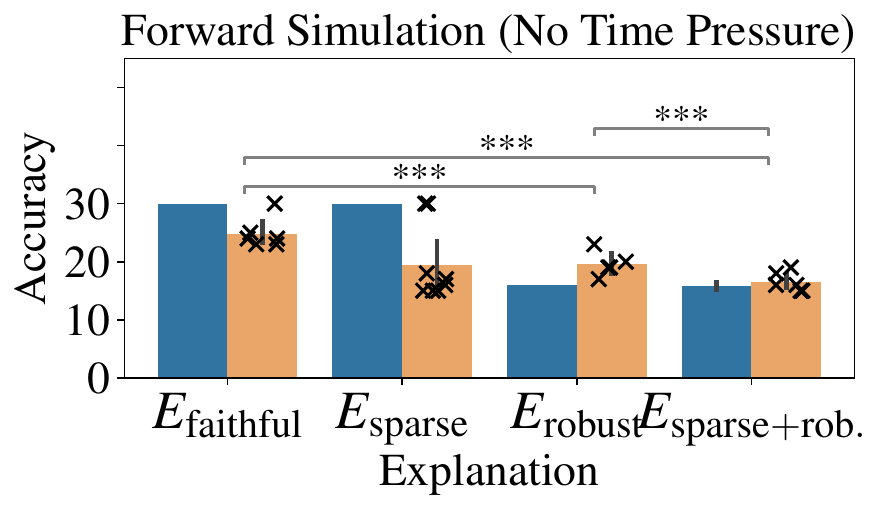}
        \caption{}
        \label{fig: human-forward-no-pressure-top}
    \end{subfigure}%
    \begin{subfigure}{0.25\linewidth}
        \includegraphics[width=1\linewidth]{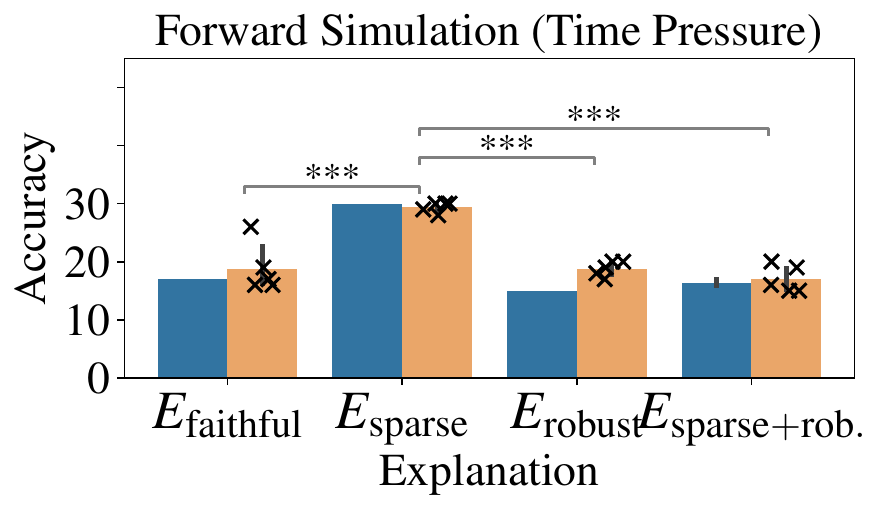}
        \caption{}
        \label{fig: human-forward-pressure-top}
    \end{subfigure}

    \caption{Task performances of participants. Each "X" is a participant. 
    Error bars are 95\% confidence intervals across participants. Significant relationships (according to one-sided Brunner-Munzel test, adjusted for multiple comparisons) are given brackets.
    We split users into two categories, by median performance within each condition; the top row are the results for all users and the bottom row is the results for users in the top $50\%$. }
    \label{fig: human-results}
\end{figure}

\paragraph{The simulation matched the performance of high-performing users.}
In \cref{tab: proxy-vs-real-relationships}, we compared the relationships observed among explanation types between our proxy users and real users. Overall, $7$ out of the $17$ effects from proxy users were verified on real users.
In \cref{fig: human-results}, there were \textit{no cases} where the proxy user performed better with one explanation over another but the real user had the opposite performance-- an indication that XAIsim2real did not capture spurious relationships between properties and tasks. Likewise, we did not observe effects on real users that were not first detected in the synthetic study. 

Based on the plot of participant performance in \cref{fig: human-results}, we believe participants fell into one of two groups: those who developed a strategy that utilized the explanation effectively and those who did not. A clear example of this is in \cref{fig: human-forward-pressure-all}, where there is a cluster of participants around $30$ (near-perfect performance) and another cluster around $15$ (random guessing). Therefore, we also report results for participants in the top $50$-th percentile within each condition. 
In \cref{tab: proxy-vs-real-relationships}, we see that among top participants, the majority of the relationships from the proxy users held ($13$ out of $17$). 
Next, we discuss specifics of why each relationship did or did not hold.

\begin{table}[h]
    \centering
    \begin{tabular}{l|l|l | l}
        Task & Proxy user & Real User (all)&  Real User (top 50\% performing)\\ \toprule
        Forbidden features 
        & $E_\text{faithful} > E_\text{sparse}$ 
            & $\boldsymbol{E_\textbf{faithful} > E_\textbf{sparse}}$\quad\ \ \ \ ($p<0.01$)
            & $\boldsymbol{E_\textbf{faithful} > E_\textbf{sparse}}$\quad\ \ \ \ ($p<0.01$)\\
        &  $E_\text{faithful} > E_\text{robust}$ 
            & $\boldsymbol{E_\textbf{faithful} > E_\textbf{robust}}$\quad\ \ \ \ (${p<0.01}$)
            & $\boldsymbol{E_\textbf{faithful} > E_\textbf{robust}}$\quad\ \ \ \ ($p<0.01$)\\
        &  $E_\text{faithful} > E_\text{sparse+rob.}$
            & $\boldsymbol{E_\textbf{faithful} > E_\textbf{sparse+rob.}}$ ($p<0.01$)
            & $\boldsymbol{E_\textbf{faithful} > E_\textbf{sparse+rob.}}$ ($p<0.01$)\\
        \midrule
        Counterfactual sim. 
        & $E_\text{robust} > E_\text{faithful}$ 
            & $\boldsymbol{E_\textbf{robust} > E_\textbf{faithful}}$\quad\quad($p=0.01$)
            & $\boldsymbol{E_\textbf{robust} > E_\textbf{faithful}}$\quad\quad  ($p<0.01$)\\
        & $E_\text{robust} > E_\text{sparse}$
            &  $\boldsymbol{E_\textbf{robust} > E_\textbf{sparse}}$\quad\quad\ \ ($p<0.01$)
            &  $\boldsymbol{E_\textbf{robust} > E_\textbf{sparse}}$\quad\quad\ \ ($p<0.01$) \\
        & $E_\text{robust} > E_\text{sparse+rob}$
            & $E_\text{robust} \approx E_\text{sparse+rob}$\quad\ \ \ ($p=\text{n.s.}$)
            & $\boldsymbol{E_\textbf{robust} > E_\textbf{sparse+rob}}$\quad($p=0.01$)\\
        & $E_\text{sparse+rob} > E_\text{faithful}$
            & $\boldsymbol{E_\textbf{sparse+rob} > E_\textbf{faithful}}$\ \  ($p=0.01$)
            & $\boldsymbol{E_\textbf{sparse+rob} > E_\textbf{faithful}}$\ \ ($p<0.01$)\\
        & $E_\text{sparse+rob} > E_\text{sparse}$
            & $E_\text{sparse+rob} \approx E_\text{sparse}$\quad \ \ \  ($p=0.02$)
            & $\boldsymbol{E_\textbf{sparse+rob} > E_\textbf{sparse}}$\quad  ($p<0.01$)\\
        \midrule
        Forward sim. 
        & $E_\text{faithful} > E_\text{robust}$
            & $E_\text{faithful} \approx E_\text{robust}$\quad\quad\ \ \  ($p=0.02$)
            & $\boldsymbol{E_\textbf{faithful} > E_\textbf{robust}}$\quad\quad\  ($p<0.01$)\\
        & $E_\text{faithful} > E_\text{sparse+rob.}$ 
            & $\boldsymbol{E_\textbf{faithful} > E_\textbf{sparse+rob.}}$\  ($p=0.01$)
            & $\boldsymbol{E_\textbf{faithful} > E_\textbf{sparse+rob.}}$\  ($p<0.01$)\\
        & $E_\text{sparse} > E_\text{robust}$ 
            & $E_\text{sparse} \approx E_\text{robust}$\quad\quad\quad \ ($p=\text{n.s}$)
            & $E_\text{sparse} \approx E_\text{robust}$\quad\quad\quad \  ($p=\text{n.s}$)\\
        & $E_\text{sparse} > E_\text{sparse+rob.}$ 
            & $\boldsymbol{E_\textbf{sparse} > E_\textbf{sparse+rob.}}$\quad  ($p=\text{n.s}$)
            & $\boldsymbol{E_\textbf{sparse} > E_\textbf{sparse+rob.}}$\quad  ($p=\text{n.s}$)\\
        \midrule
        Forward sim. (tp)
        & $E_\text{sparse} > E_\text{faithful}$
            & $E_\text{sparse} \approx E_\text{faithful}$\quad\quad\quad ($p=\text{n.s}$)
            & $\boldsymbol{E_\textbf{sparse} > E_\textbf{faithful}}$\quad\quad\    ($p<0.01$) \\
        & $E_\text{sparse} > E_\text{robust}$ 
            & $E_\text{sparse} \approx E_\text{robust}$\quad\quad\quad \  ($p=\text{n.s}$)
            & $\boldsymbol{E_\textbf{sparse} > E_\textbf{robust}}$\quad\quad\ \ \  ($p<0.01$)\\
        & $E_\text{sparse} > E_\text{sparse+rob.}$ 
            & $E_\text{sparse} \approx E_\text{sparse+rob.}$\quad\ \ ($p=\text{n.s}$)
            & $\boldsymbol{E_\textbf{sparse} > E_\textbf{sparse+rob.}}$\quad($p<0.01$)\\
        & $E_\text{faithful} > E_\text{robust}$ 
            & $E_\text{faithful} \approx E_\text{robust}$\quad\quad\ \ \ \ ($p=\text{n.s}$)
            & $E_\text{faithful} \approx E_\text{robust}$\quad\quad\quad ($p=\text{n.s}$)\\
        & $E_\text{faithful} > E_\text{sparse+rob.}$ 
            &$E_\text{faithful} \approx E_\text{sparse+rob.}$\quad \ ($p=\text{n.s}$)
            &$E_\text{faithful} \approx E_\text{sparse+rob.}$\quad \  ($p=\text{n.s}$)\\
    \end{tabular}
    \caption{Validation of proxy user relationships on real user data. Each row is a hypothesized relationship from \cref{sec: simulated-user-results}. Real user results are from the Brunner-Munzel test, where the first column is the check on all participants and the second is on the top 50\% of participants. 
    }
    \label{tab: proxy-vs-real-relationships}
\end{table}

\paragraph{On the forbidden features task, faithful explanations were best.} This was the task where real user performance most closely followed that of the proxies, likely because it was the easiest task.
$E_\text{faithful}$ was best among all pairwise comparisons against $E_\text{robust}$ ($p<0.01$), $E_\text{sparse} (p<0.01)$, and $E_\text{sparse+rob.} (p<0.01)$. 
The real users performed better than the proxy users due to technicalities of the training instances. 
The proxy user (a decision tree) learned a threshold that perfectly separated compliant and non-compliant cases, but this threshold was as large as possible. In our qualitative analysis of real user strategies, we found that humans tended to learn to threshold at exactly zero. The test instances included cases where the threshold needed to be $0$-- the real users did better on these instances than the proxy. This distinction wouldn't have shown up if we had either (1) regularized the proxy \textit{learning model} so that it tended to learn thresholds near $0$ or (2) included training instances where a smaller threshold was needed for perfect separation.

\paragraph{On the counterfactual simulation task, only top-performing users performed as well as expected with sparse and sparse+robust explanations.}
When considering all users, $E_\text{robust}$ was best only when tested against $E_\text{sparse}$ ($p<0.01$). When considering top users, all expected relationships held, except that $E_\text{robust}$ was preferred over $E_\text{sparse+rob}$ only for the top-performing users.

\paragraph{On the forward prediction task without time pressure, only top-performing users performed as expected with sparse and faithful explanations.} 
The forward prediction task required the most mental math of all of the tasks, which resulted in noisier performance among our participants. Without time pressure, the expected relationships held among the top users. With time pressure, $E_\text{faithful}$ was worse than $E_\text{sparse}$ ($p<0.01$), but it did not perform better than $E_\text{robust}$ nor $E_\text{sparse+rob.}$. This suggests that time pressure affects real users more than the proxies.

\begin{figure}[h]
    \centering
    \includegraphics[width=0.5\linewidth]{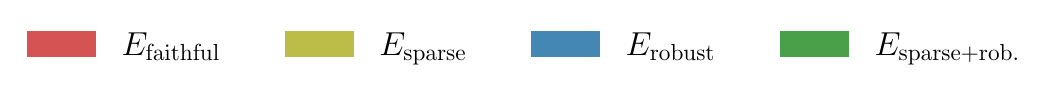}
    \begin{subfigure}{0.33\linewidth}
        \includegraphics[width=1\linewidth]{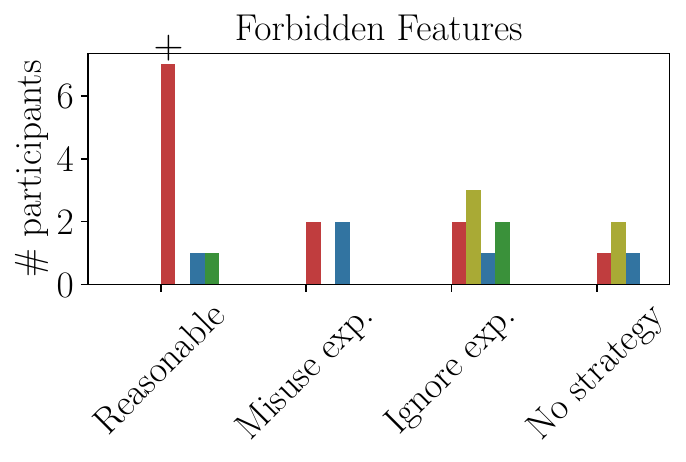}
        \caption{}
        \label{fig: qualitative-forbidden-features}
    \end{subfigure}%
    \begin{subfigure}{0.33\linewidth}
        \includegraphics[width=1\linewidth]{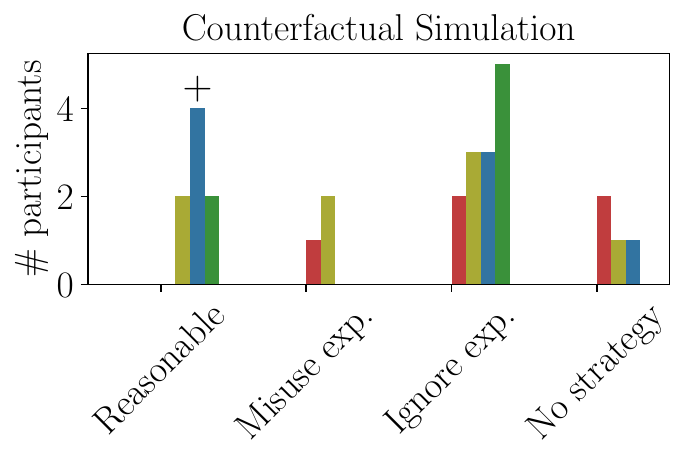}
        \caption{}
        \label{fig: qualitative-counterfactual-simulation}
    \end{subfigure}%
    \begin{subfigure}{0.33\linewidth}
        \includegraphics[width=1\linewidth]{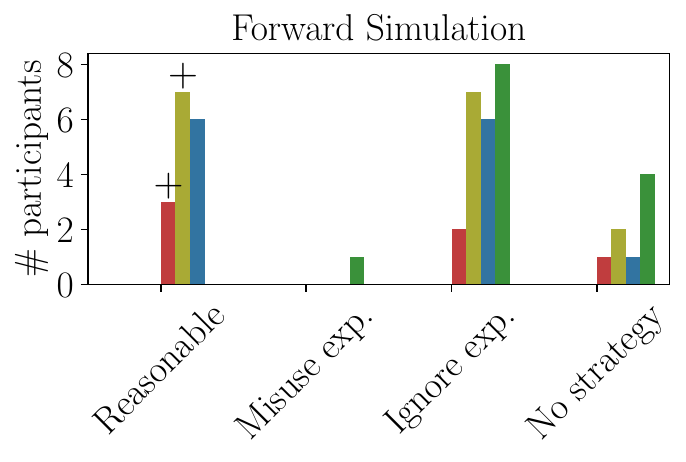}
        \caption{}
        \label{fig: qualitative-forward-simulation}
    \end{subfigure}%
    \caption{Break down of strategy codes by group and explanation. More participants were able to form reasonable strategies when helpful properties (marked with a `+') were present. 
    }
    \label{fig: qualitative}
\end{figure}

\paragraph{Helpful properties lead participants to find reasonable strategies more often.}
In \cref{fig: qualitative}, we show the breakdown of participant strategies in each task. Below, we elaborate on why some explanation types led to misuse and disuse of the explanations. The full table of codes and groupings is in \cref{appendix: codes}.

\paragraph{Some participants with \textbf{no strategy} found the explanations unhelpful.}
Several participants (15 out of 156) reported having no strategy upfront (e.g. \participant{To be completely honest, I don't have a strategy.}), especially when the helpful property was not present. Four participants doubted the validity of the explanation when helpful properties were not present. For example, one participant with sparse explanations on forbidden features said: \participant{I would like to rely on the researcher's judgment, but I'm not sure I can since it seems to think Glow was 0 even in the non-compliant examples.}

\paragraph{Participants who \textbf{misused} the explanations had misconceptions about explanation or the task.}
In forward simulation, one participant thought that since explanations described what features mattered in deciding if an alien is healthy, the inputs should match the explanation: \participant{See if the measurements for the important categories match what the helpful information says.} In forbidden features, four participants checked whether the attribution for the forbidden feature (``glow'') was the smallest, rather than checking whether it was used at all: \participant{I look to see if there are any other measurements or information that exceeds (or is less than 0) the score of the glow}. Finally, three participants applied strategies that would be reasonable in forward simulation (weighting the attributions by the input, summing, and thresholding) to the one that they were assigned, possibly because they formed a mental model on how explanations could be used independent of the task.

\paragraph{Participants who \textbf{ignored} the explanations tried to reverse engineer the behavior of the underlying AI function instead.} Because the behavior of the function is more complicated than one could learn over ten training instances, 16 participant strategies were based on a set of (incorrect) rules derived from the examples: \participant{glow is more often illegally used when it measures at 0 and 40, and sometimes at 30, possibly when the temp. is high or the brain activity is low.}. Nine participants based their decisions solely on the largest or smallest input features (and not explanations): \participant{Looking at the highest number in measurement.} 

Surprisingly, in the counterfactual simulation task, participants with robust explanations learned high-performance strategies that did not rely on explanations.
Like with the other tasks, these strategies reverse-engineered the AI function from the training examples: \participant{Pulse rate increase and skin moisture increase means it [risk score] will go down, likewise a decrease causes an increase [in risk score]. Hearing score increase increases risk}. What is surprising here is that participants in the $E_\text{sparse}$ and $E_\text{faithful}$ conditions, who were also capable of ignoring the explanations, did not do so as often. It may be that the robust explanations were easier to ignore because they did not change, which in turn, allowed participants to focus attention on how other inputs related to the best decision. 

\section{Discussion}
In this paper, we introduced XAIsim2real, a novel pipeline for simulating user studies where explanation properties are the design space.
In our validation study, we made design choices (e.g. by choosing specific and simple forms of AI decision functions) that produced strong signals in simulation, so that we could demonstrate that \textit{XAIsim2real successfully identifies property trade-offs when they do exist}. 
Our successful validation means that other researchers and practitioners may use the pipeline to explore the impact of properties in their own settings. 
In other settings, XAIsim2real may reveal that there is not a trade-off among properties, meaning there is no choice of explanation to be made; this is still an important finding to have prior to running a user study. 

Below, we discuss the insights from our validation study and their implications for XAI; we also describe how other researchers and practitioners can extend XAIsim2real to human-AI interaction paradigms beyond the one studied in this paper.


\subsection{How do our findings compare with existing studies? Our findings support observations from prior work by formally linking different explanation properties to user performance on different tasks (RQ2). We also shed new light on results from existing studies.}
We built on recent literature that has suggested that different properties may be required for different tasks \citep{jesus2021can}, by running a user study where explanation properties are the independent variable and the outcome is task-performance. 
The property-task relationships we found in our user studies validate what has been suggested in prior literature. Specifically,  \citet{lertvittayakumjorn2019humanGroundedText} and \citet{nguyen2018comparing} found that local fidelity of the explanation is important for forward simulation, this is supported by the results from our user study, in the setting without time pressure. On the other hand, \citet{poursabzi2021manipulating} found that compactness (equal to our sparsity) also mattered in their forward simulation task. By using time pressure as a means to constrain the user's cognitive budget, we found that sparsity mattered more when time pressure was applied. Our findings may explain why existing studies on forward simulation seemingly disagree on the importance of compactness for this task.

Our qualitative analysis also found that participants were able to quickly identify situations where the explanations were unhelpful for the task, when the explanations were robust. This aligns with findings in \citet{jesus2021can}, in which experts preferred LIME explanations the least because they showed the ``least diversity'' (i.e. were too robust). However, we also found that robustness could be useful for the human decision-maker for the very same reason. That is, participants who were quick to notice robust explanations were unhelpful were also more likely to find alternate decision-making strategies that did not rely on the explanations. 

\subsection{To what extent can computational proxies simulate real users? When our computational simulations showed strong patterns, we found them in studies with real users (RQ1). Patterns in real user studies can teach us how to improve our proxy user models.}
Unsurprisingly, we observed higher variance in performance among real users compared to proxy users. This suggests that subtle inter-property differences identified by XAIsim2real may not be evident with real users or require larger sample sizes to detect. In the counterfactual simulation task, robust explanations slightly outperformed robust+sparse ones for proxy users. While real users showed a similar trend (see \cref{fig: human-counterfactual-sim-top}), the difference was not statistically significant.
On the other hand, when XAIsim2real revealed notable differences between properties, we easily observed them in real users, especially among top performers. It is promising that XAIsim2real can predict performance differences in simulation, even with a general proxy user model that was not specifically designed for each task. 

Nonetheless, it is interesting to consider how we might improve the proxy user, and what these choices might teach us about the decision-making of real users. 
For example, in XAIsim2real, we proposed a novel proxy user model that consists of a general memory model (how do humans process information) and a learning model (how do humans learn from experience). This way of constructing a proxy user has not been studied before in literature~\citep{chen2022useCaseSimulations}, and it provides explicit ways to evaluate our assumptions about real users. For example, in our case-study, we assumed that real users were less willing to perform math under time-pressure, and we captured this assumption with the memory model by feeding the proxy only partially-computed inputs. 

We also observed behaviors in the real users that could be incorporated into our memory and learning models. 
For example, in our qualitative analysis, one of our codes found that users hyper-focused on extreme values-- a type of extremity bias. We could encode this in our memory model by masking values that are not in the extremes. 
We also observed that participants in our study tend to fall into two groups-- those who figured out how to use the explanation from training and those who didn't. This implies that we could consider two types of learning models for our proxy user, one for each type of participant. 

In the future, the selection of the proxy could be treated as an iterative process. Data from a small-scale pilot study can be used to improve the proxy user, prior to conducting a full study. For example, using XAIsim2real, one can tune the hyper-parameters of the proxy user model (e.g. the depth of the decision tree) to fit the real user performance in the pilot data.

Finally, in our validation study, we trained separate proxy models for each task, rather than a single proxy to perform multiple tasks. This approach was not computationally burdensome, as we utilized ``lightweight'' proxy models that required only ten training instances per task, minimizing the computational load. While generalized models like large language models (LLMs) could handle multiple tasks, they require caution. LLMs often make implicit assumptions about users' \textit{memory} and \textit{learning}, such as their ability to retain and apply complex explanations, which may differ from actual user behavior, potentially leading to mismatches between proxy and real performance.

\subsection{Does XAIsim2real generalize to other human-AI decision-making paradigms? By changing the memory model or the learning model in our proxy user, we can incorporate a broader range of assumptions about real users, paradigms of human-AI interactions, as well as capture other important outcomes (RQ3).}
Our case-study involves a specific format of human-AI interaction: the explanation is available for every decision, people are not instructed how to use the explanation to do the task, and outcomes are measured as task-performance. Below, we discuss how XAIsim2real may (or may not) extend to human-AI interaction settings beyond the one we considered. 

In our case study, we've demonstrated that XAIsim2real is a useful framework for evaluative tasks, where the correct human decision is known, and the user study evaluates whether humans can arrive at the correct decision with the explanation (i.e. the outcome we measure is accuracy). Although evaluative tasks are commonly studied in user studies \citet{rong2023surveyUserStudies}, there are many other outcomes in human-AI interactions that are important to study. For example, we may wish to understand the impact of explanation properties on the perceived usefulness or the user satisfaction of an explanation. We can potentially collect real user data for these outcomes with a pilot study, and train a proxy user to predict satisfaction as reported by pilot users. Outcomes like time-to-decision for a real user may be approximated by the number of computations a user needs to do to arrive at a decision, and this we can capture as the depth of the decision tree for our learning model.

In this paper, we only considered situations where the explanation was always available, and did not consider outcomes such as over-reliance. Adapting XAIsim2real to studies where the explanation is available only part of the time could be as straightforward as replacing the part of the proxy's input that would normally include the explanation with ``0''s. In this case, over-reliance could be formalized as the difference between the proxy's performance on test cases with and without the explanation. That said, our work to understand how properties affect user performance is already related to recent work in reducing over-reliance. For example, \citet{lai2023selective} proposed to reduce over-reliance by inferring and \textit{selecting} explanations that are relevant to the user's decision task. In this case, we can use XAIsim2real to see which properties are needed for each task and can use this knowledge to select explanations for each user. 

In our case-study, participants had no prior knowledge and were told explicitly how to use explanations to make decisions. We can make different choices in both cases and correspondingly capture them by changing our learning model. We can represent the user's prior knowledge of the task using a Bayesian prior on the weights of our proxy user model; we can also represent the user's prior knowledge as a specific way to initialize the proxy model's weights for optimization. 

Finally, we note that although our case study focused on the impact of properties on user performance, XAIsim2real can be used to choose between explanations without considering properties. For example, just as in \citet{chen2022useCaseSimulations}, we can use XAIsim2real to directly compare between explanations generated by LIME and SHAP by skipping the step where we optimize the explanations for properties.

\subsection{Future work: directly optimizing explanations for properties when the AI decision function is complex or unknown.}
In our pipeline, we used our knowledge of the AI's decision function in order to directly produce explanations that were optimal for different properties (i.e. because we know which feature $f_\mathrm{sparse}$ relies on for the output, we know that the most faithful explanation is one that highlights this feature). However, in real-life, AI decision functions will be black-boxes or will be highly complex, and thus analytic optimization is no longer possible. We have considered one approach to optimizing explanations for properties using an unknown function: we randomly sampled the feature attributions in our explanations and filtered for samples that scored the highest for a desired property. However, we found this method to be ineffective at finding a set of explanations that have diverse types of properties (e.g. the randomly sampled explanations tended to be similarly faithful and robust). An interesting line of future work can be to find scalable and efficient ways to directly optimize explanations for desired properties, when the underlying AI function is complex or unknown.

\section{Conclusion}
In this paper, we introduced XAIsim2real, a framework that leverages simulation to predict how explanation properties impact task-performance, then uses insights from simulations to design and analyze real user studies that test these predicted relationships.
We argue that XAIsim2real offers a scalable way to explore which explanation properties are needed to help a user perform a task in a human-AI decision-making setting. We demonstrated the utility of XAIsim2real in a validation study, and we highlighted future work that can extend the computational models in XAIsim2real to incorporate a broader range of human-AI decision-making settings. 

\bibliographystyle{ACM-Reference-Format}
\bibliography{references}

\newpage
\appendix

\section{Experimental Details for Simulated User Studies}
\label{appendix: simulated-study-details}

\subsection{AI functions}
\label{appendix: ai-functions}
\textbf{Function $f_\text{sparse}$.}
This is a binary classification function that takes three-dimensional inputs $\mathbf{x} \in \mathbb{R}^3$ and returns a binary output $y = f_\text{sparse}(\mathbf{x}) \in \{0, 1\}$. 
This function relies on only one input feature at a time to make a decision-- thus making it ``sparse.''

\begin{equation}
\label{eq: box-function}
    f_\text{sparse}(\mathbf{x}) = \begin{cases}
        \ind{\mathbf{x}_2 > 0.5}, &  \mathbf{x}_3 \le 0.25\\
        \ind{\mathbf{x}_1 > 0.5}, &  0.25 < \mathbf{x}_3 \le 0.5\\
        \ind{\mathbf{x}_2 > 0.5}, &  0.5 < \mathbf{x}_3 \le 0.75\\
        \ind{\mathbf{x}_1 > 0.5}, &  0.75 < \mathbf{x}_3,
        \end{cases}
\end{equation}
where $\mathbb{I}(\mathrm{condition}(x))$ is the indicator function that returns 1 if $\mathrm{condition}(x)$ is true, 0 otherwise.
In \cref{eq: box-function}, the binary classification depends on either the first $\mathbf{x}_1$ or second $\mathbf{x}_2$ input feature. The third feature $\mathbf{x}_3$ decides which of the two is used. 

For example, suppose an input $\mathbf{x} = [0.6, 0.4, 0.3]$ is given. We aim to compute the output $y = f_\text{sparse}(\mathbf{x})$. According to the conditions in \eqref{eq: box-function}, we first check the value of the third feature $\mathbf{x}_3 = 0.3$ to determine which case to use.
Since $0.25 < \mathbf{x}_3 \leq 0.5$, the second case applies: $f_\text{sparse}(\mathbf{x}) = \mathbb{I}(\mathbf{x}_1 > 0.5).$
Next, we evaluate the condition on $\mathbf{x}_1 = 0.6$. Since $0.6 > 0.5$, the indicator function returns 1:
$\mathbb{I}(\mathbf{x}_1 > 0.5) = 1.$
Thus, for the input $\mathbf{x} = [0.6, 0.4, 0.3]$, the function outputs:
$y = f_\text{sparse}(\mathbf{x}) = 1$.

\textbf{Function $f_\text{trend+wiggle}$.}
This is a regression function that has a long-term linear trend with a locally periodic trend. It takes five-dimensional inputs $\mathbf{x} \in \mathbb{R}^5$ and returns a real-valued output $f_\text{trend+wiggle}(\mathbf{x}) = y \in \mathbb{R}$. 
Explanations that are locally faithful to this function are not robust, because they capture the local oscillations. 
Explanations that are more ``globally'' faithful to this function are robust, because they capture the long-term linear trend. 

\begin{equation}
    f_\text{trend+wiggle}(x) = \sum\limits_{d}^D 5 \cdot \sin{20 \mathbf{x}_d} + \mathbf{w}_d \mathbf{x}_d
\end{equation}
where $\mathbf{w} = [20, -1, -20, 1]$. In plain words, the long-term linear trend will be either upwards or downwards. Some of the linear trend's weights are smaller so that a sparse explanation will not do too poorly-- this relates to how we control for baseline properties of explanations. 

\textbf{Function $f_\text{wiggle}$.}
This is a binary classifier whose dependence on the inputs is highly local; depending on where the input is located, different features have very different effects on the output. Explanations that are faithful to this function are not robust nor sparse. 
It takes ten-dimensional inputs $\mathbf{x} \in \mathbb{R}^10$ and returns a binary output $f_\text{wiggle}(\mathbf{x}) = y \in \{0, 1\}$. 
In this paper, for simplicity, we choose to work with a piecewise linear function, but we note that other functions, such as splines, can also satisfy the ``highly local'' quality. 

\begin{equation}
\label{eq: piecewise-function}
    f_\text{wiggle}(x) = \begin{cases}
        \ind{\mathbf{x}^\top W_{1:} > 0}, &  \mathbf{x}_1 \le 0.25\\
        \ind{\mathbf{x}^\top W_{2:}> 0} , &  0.25 < \mathbf{x}_1 \le 0.5\\
        \ind{\mathbf{x}^\top W_{3:}> 0} , &  0.5 < \mathbf{x}_1 \le 0.75\\
        \ind{\mathbf{x}^\top W_{4:}> 0} , &  0.75 < \mathbf{x}_1,
        \end{cases}
\end{equation}
where $W_{i:}$ refers to the $i$-th row of the weight matrix, defined as 
$$ W = 
\begin{pmatrix}
    0 &  1 & -1 & 0 & 1 & -0.1 & 0.1 & -0.1 & 0.1 & -0.1 & -0.7 \\
    0 & -0.8 & -0.2 & 0.2 & 0.1 & -0.9 & -0.1 & -0.1 & 0.1 & -0.2 & 1\\
    0 & -0.8 & -0.2 & 0 & 0.1 & -0.9 & -0.1 & -0.1 & 0.1 & -0.2 & 1\\
    0 & -0.05 & 1 & -0.8 & -0.1 & 0.1 & 0.9 & -0.2 & 0.1 & 0.8 & -1.
\end{pmatrix}
$$

For example, suppose an input $\mathbf{x} = [0.3, 0.7, -0.4, 0.1, 0.5, -0.2, 0.3, -0.1, 0.4, -0.5]$ is given. To compute the output $y = f_\text{wiggle}(\mathbf{x})$, we first check the value of $\mathbf{x}_1 = 0.3$. Since $0.25 < \mathbf{x}_1 \leq 0.5$, the second case in \eqref{eq: piecewise-function} applies, meaning that $f_\text{wiggle}(\mathbf{x}) = \mathbb{I}(\mathbf{x}^\top W_{2:} > 0)$, where $W_{2:}$ is the second row of the weight matrix $W$. 
Next, we compute the dot product $\mathbf{x}^\top W_{2:}$, excluding the first element of $\mathbf{x}$ as it multiplies by $0$. The dot product is:
$W_{2:}^\top \mathbf{x} = 0.89$.
Since $0.89 > 0$, the indicator function $\mathbb{I}(\mathbf{x}^\top W_{2:} > 0)$ returns 1. Therefore, for this input, the output is $y = f_\text{wiggle}(\mathbf{x}) = 1$. 

\subsection{Optimization methods for property-optimized explanations}
\label{appendix: explanation-optimization}

\paragraph{Optimizing on a per-function basis.}
As stated in the main body, we optimized explanations to properties on a function-by-function basis. The task of optimizing explanations to properties for general functions is future work. 

We assume access to a dataset of inputs $X \in \mathbb{R}^{N \times D}$, with $N$ points each of $D$ dimension. The corresponding set of property-optimized feature attributions will also be $W \in \mathbb{R}^{N \times D}$. Let $X_n$ and $W_n$ refer to the input and explanation for the $n$-th point, respectively.

The optimization objective for each property is based directly on the equations from \cref{sec: components-properties} applied to the optimization problem in \cref{eq: optimization_problem}:
\begin{itemize}
    \item $\mathcal{L}_\text{faithful}(W) = \sum_{n=1}^N \text{fidelity\_classification\_loss}(X_n, W_n, p) $
    \item $\mathcal{L}_\text{sparse} = \sum_{n=1}^N \text{fidelity\_classification\_loss}(X_n, W_n, p) + \text{sparsity\_loss}(W_n)$
    \item $\mathcal{L}_\text{robust} = \sum_{n=1}^N \text{fidelity\_classification\_loss}(X_n, W_n, p) + \text{robustness\_loss}(X_n, W_n, r)$
    \item $\mathcal{L}_\text{sparse+rob} = \sum_{n=1}^N \text{fidelity\_classification\_loss}(X_n, W_n, p) + \text{sparsity\_loss}(W_n) + \text{robustness\_loss}(X_n, W_n, r)$
\end{itemize}

Note that the purpose of finding the property-optimized explanation is to run a user-study where the properties are the manipulated variable. As such, our goal is not to find the \textit{global optima}, but rather, to find explanations that are optimized enough along each property axis to appear distinct from one another.

Let $W^*$ refer to our optimized explanation. 
For each property-function pair, we optimized for $W^*$ either by (1) sampling or (2) used hardcoded knowledge of the underlying AI function. In \cref{tab: optimization}, we describe which strategy we applied to each pair. We expand on how we implemented each strategy here: 
\begin{itemize}
    \item \textbf{Sample global explanations.} 
    We took $S$ total samples of $W$ and kept the one with the lowest loss: 
    \begin{equation}
    \label{eq: optimize-by-sampling}
    W^* = \min_{W^s} \mathcal{L}(W^s),    
    \end{equation}
    
    where $\mathcal{L}$ is the loss function and $W^s$ is one of $S$ total explanation samples. 

    This strategy involves sampling global explanations because they will always evaluate to perfect robustness. 
    We sample a single $W^s$ as follows. Since we are interested in \textit{global} explanations, all rows of $W$ should be the same. In other words, the feature attribution should not change across instances. So, to sample the matrix of feature attributions, we first sample a single row, $$\mathbf{w} \sim \text{Uniform}(0, 1)^D.$$ Then, we form the matrix by duplicating the vector $\mathbf{w}$ for all $N$ rows: $$W^s = \begin{bmatrix} \mathbf{w}\  \mathbf{w}\  \ldots \ \mathbf{w} \end{bmatrix}^\top \in \mathbb{R}^{N \times D}.$$

    \item \textbf{Sample global explanations (only 2).}
    Exactly as in \cref{eq: optimize-by-sampling}, this strategy samples the feature attributions and keeps the one that evaluates to the lowest loss. However, we alter the sampling scheme for a single $W^s$ to ensure each of the samples is \textit{sparse} in addition to being robust. 
    First, we sample a single row from a uniform as before, $$\mathbf{w} \sim \text{Uniform}(0, 1)^D.$$ 
    Then, we sample two of the dimensions to keep and mask the remaining to be zero (this is the step that ensures sparsity): $$\tilde{\mathbf{w}} = \mathbf{w} \odot \mathbf{m},$$ where $\mathbf{m} \in \{0, 1\}^D$ is a binary mask with exactly two entries equal to 1. 
    Finally, we form the matrix by duplicating the vector $\mathbf{w}$ for all $N$ rows: $$W^s = \begin{bmatrix} \mathbf{w}\  \mathbf{w}\  \ldots \ \mathbf{w} \end{bmatrix}^\top \in \mathbb{R}^{N \times D}.$$
    
    \item \textbf{Ground-truth weights.}
    This strategy assumes the underlying AI function is linear around each input-- an assumption that holds for both $f_\text{sparse}$ and $f_\text{wiggle}$. For this strategy, we return the ground-truth weights of the linear AI function for each input. 
    Let $\mathbf{w}^*_n$ refer to the ground-truth weights of the AI function $f$ at the $n$-th point, so that $f(X_n) = \mathbf{w}_n^* \mathbf{x}_n$. Then the feature attributions are : $$W^* = \begin{bmatrix}
        \mathbf{w}^*_1\\
        \mathbf{w}^*_2\\
        \vdots\\
        \mathbf{w}^*_N\\
    \end{bmatrix} \in \mathbb{R}^{N \times D}.$$ 
    \item \textbf{Fit locally faithful line.}    
    When the underlying AI function is not linear (as is the case for $f_{\text{trend+wiggle}}$), we approximate the function locally with a linear model. The explanation is the weights of this locally linear approximation. 
    
    For each input point $\mathbf{x}_n$, we sample a set of neighboring points $\mathcal{N}(\mathbf{x}_n) = \{\mathbf{x}_n^{(1)}, \mathbf{x}_n^{(2)}, \dots, \mathbf{x}_n^{(S)}\}$, where $S$ is the number of sampled points around $\mathbf{x}_n$.

    We then fit a linear model to the values of the function evaluated at these points, $f(\mathbf{x}_n^{(s)})$, for $s = 1, 2, \dots, S$. Let $\hat{\mathbf{w}}_n$ denote the coefficients of the fitted linear model at $\mathbf{x}_n$, obtained via least-squares regression:
    \[
    \hat{\mathbf{w}}_n = \arg\min_{\mathbf{w}} \sum_{s=1}^S \left( f(\mathbf{x}_n^{(s)}) - \mathbf{w}^\top \mathbf{x}_n^{(s)} \right)^2.
    \]
    
    The matrix of approximated feature attributions $W^*$ is then formed as:
    \[
    W^* = \begin{bmatrix}
        \hat{\mathbf{w}}_1 \\
        \hat{\mathbf{w}}_2 \\
        \vdots \\
        \hat{\mathbf{w}}_N
    \end{bmatrix} \in \mathbb{R}^{N \times D},
    \]
    where $\hat{\mathbf{w}}_n$ represents the linear approximation of $f$ around $\mathbf{x}_n$, and $N$ is the total number of input points.
    
    \item \textbf{Top 2 from faithful.} This strategy takes an explanation that has already been optimized for faithfulness and then makes it sparse by returning only the largest two attributions per input. 
    Let $W \in \mathbb{R}^{N \times D}$ refer to the matrix of feature attributions that has already been optimized for faithfulness, where each row $\mathbf{w}_n \in \mathbb{R}^D$ represents the feature attributions for input $\mathbf{x}_n$. 
    
    To make the explanation sparse, we retain only the two largest (in magnitude) attributions for each input and set the rest to zero.
    Let $\mathbf{w}_n^{\text{sparse}}$ denote the sparse version of the attribution vector for input $\mathbf{x}_n$. We define this as follows:
    \[
    \mathbf{w}_n^{\text{sparse}}[i] = \begin{cases}
    \mathbf{w}_n[i], & \text{if } i \in \mathcal{I}_n \\
    0, & \text{otherwise}
    \end{cases}
    \]
    where $\mathcal{I}_n$ is the set of indices corresponding to the two largest (in magnitude) elements of $\mathbf{w}_n$.
    
    The matrix of sparse attributions $W^* \in \mathbb{R}^{N \times D}$ is then formed as:
    \[
    W^* = \begin{bmatrix}
        \mathbf{w}_1^{\text{sparse}} \\
        \mathbf{w}_2^{\text{sparse}} \\
        \vdots \\
        \mathbf{w}_N^{\text{sparse}}
    \end{bmatrix}.
    \]
    
\end{itemize}

\begin{table}[h]
    \centering
    \begin{tabular}{c|c|c|c | c}
         & Faithful & Sparse & Robust & Sparse and Robust \\
         \toprule
        $f_\text{sparse}$ & Ground-truth weights$^\dagger$& Ground-truth weights & Sample global explanations & Sample global explanations (only 2)\\
        $f_\text{trend+wiggle}$ & Fit locally faithful line &  Top 2 from faithful$^\dagger$ & Ground-truth weights & Sample global explanations (only 2)\\
        $f_\text{wiggle}$ & Ground-truth weights&  Top 2 from faithful$^\dagger$ & Sample global explanations & Sample global explanations (only 2)\\ 
    \end{tabular}
    \caption{Optimization strategies for each property-function pair. The $\dagger$ on some of the strategies indicates that we added a post-optimization step to control for confounding of certain properties. For example, because faithful and sparse explanations are equivalent for $f_\text{sparse}$, we needed a way to make the faithful explanation ``less sparse'' so sparsity was not a confounder.} 
    \label{tab: optimization}
\end{table}

\paragraph{Controlling for confounding}
Since our property-optimized explanations are used as independent variables, we want to minimize the confounding of properties that were not manipulated within each explanation type. For example, we want faithful explanations $E_\text{faithful}$ and sparse explanations $E_\text{sparse}$ to evaluate to the same level of robustness, so that robustness is not a confounding factor. 

To control for confounding, we made manual adjustments to the three optimization process in \cref{tab: optimization} marked with $\dagger$: 
\begin{itemize}
    \item We had to make faithful explanations for $f_\text{sparse}$ less sparse. This is because $f_\text{sparse}$ is defined in such a way that sparse explanations are also perfectly faithful. 

    To reduce sparsity, we modified the explanations by adjusting attributions that were zero. Specifically, we added $0.6$ to any attribution that originally had a value of zero. If there were two zero attributions in the same row, we subtracted $0.6$ from one of the zeros and added $0.6$ to another, so that we preserved the overall faithfulness.
    
    Let $\mathbf{w}_n \in \mathbb{R}^D$ be the attribution vector for the input $\mathbf{x}_n$, and let the modified attribution vector be denoted as $\mathbf{w}_n^{\text{mod}}$. The modification is applied as follows:
    
    \[
    \mathbf{w}_n^{\text{mod}}[i] = \begin{cases}
    \mathbf{w}_n[i] + 0.6, & \text{if } \mathbf{w}_n[i] = 0 \text{ and } \sum_{j=1}^D \ind{\mathbf{w}_n[j] = 0} = 1 \\
    \mathbf{w}_n[i] - 0.6, & \text{if } \mathbf{w}_n[i] = 0 \text{ and } \sum_{j=1}^D \ind{\mathbf{w}_n[j] = 0} = 2 \\
    \mathbf{w}_n[i], & \text{otherwise}
    \end{cases}
    \]
    
    Here, $\mathbb{I}(\cdot)$ is the indicator function, which counts how many zeros are present in the attribution vector. This ensures that explanations become less sparse while maintaining faithfulness for $f_\text{sparse}$.

    \item We had to make sparse explanations for $f_\text{trend+wiggle}$ and $f_\text{wiggle}$ less robust.  To do so, we scaled random rows of $W$ by a factor of 4.     
    By increasing the magnitude of the attributions in some rows, the difference between the perturbed attributions and the original attributions becomes larger in response to even small changes in the input. This makes the attributions more sensitive to variations, thereby reducing the overall robustness of the explanation.
\end{itemize}

\paragraph{Optimization results.}
In \cref{tab: optimization_results}, we verify that our optimization was successful. We base success on two checks. First, \textit{we optimized the explanations to their respective properties}. In \cref{tab: optimization_results}, each explanation type scores best on the property for which it was optimized. For example, $E_\text{robust}$ and $E_\text{sparse+robust}$ evaluate best on robustness when compared to all other explanation types. 

Second, \textit{we controlled for confounding}. In \cref{tab: optimization_results}, explanations score similarly on properties for which they are not being optimized. For example, all of the non-sparse explanations ($E_\text{faithful}$ and $E_\text{robust}$) evaluate to similar levels of sparsity. 

\begin{table}[h]
\centering
\begin{tabular}{llllll}
\toprule
 &  & $E_\text{faithful}$ & $E_\text{sparse}$ & $E_\text{robust}$ & $E_\text{sparse+rob.}$ \\
\midrule
\multirow[t]{5}{*}{$f_\text{sparse}$} & Faith(r=0) & $\boldsymbol{0.07 \pm 0.02}$ & $\boldsymbol{0.04 \pm 0.01}$ & $0.34 \pm 0.03$ & $0.40 \pm 0.03$ \\
 & Faith.(r=0.01) & $0.62 \pm 0.00$ & $\boldsymbol{0.55 \pm 0.01}$ & $0.63 \pm 0.01$ & $0.66 \pm 0.01$ \\
 & Faith.(r=0.05) & $0.63 \pm 0.01$ & $\boldsymbol{0.55 \pm 0.01}$ & $0.63 \pm 0.01$ & $0.66 \pm 0.01$ \\
 & Sparsity & $3.91 \pm 0.02$ & $\boldsymbol{2.0 \pm 0.0}$ & $4.00 \pm 0.00$ & $\boldsymbol{2.0 \pm 0.0}$ \\
 & Local Stability ($r=0.1$) & $45.80 \pm 17.29$ & $59.83 \pm 2.85$ & $\boldsymbol{0.0 \pm 0.0}$ & $\boldsymbol{0.0 \pm 0.0}$ \\
\cline{1-6}
\multirow[t]{5}{*}{$f_\text{wiggle}$} & Faith.(r=0) & $\boldsymbol{0.05 \pm 0.01}$ & $0.45 \pm 0.03$ & $0.43 \pm 0.03$ & $0.41 \pm 0.03$ \\
 & Faith.(r=0.01) & $\boldsymbol{0.5 \pm 0.02}$ & $0.88 \pm 0.06$ & $0.95 \pm 0.10$ & $0.67 \pm 0.01$ \\
 & Faith.(r=0.05) & $\boldsymbol{0.53 \pm 0.03}$ & $0.90 \pm 0.05$ & $0.94 \pm 0.09$ & $0.67 \pm 0.01$ \\
 & Sparsity & $9.46 \pm 0.03$ & $\boldsymbol{2.0 \pm 0.0}$ & $9.00 \pm 0.00$ & $\boldsymbol{2.0 \pm 0.0}$ \\
 & Local stability ($r=2$) & $15.32 \pm 0.31$ & $12.86 \pm 0.24$ & $\boldsymbol{0.0 \pm 0.0}$ & $\boldsymbol{0.0 \pm 0.0}$ \\
\cline{1-6}
\multirow[t]{3}{*}{$f_\text{trend+wiggle}$} & Faith.(r=0.05) & $\boldsymbol{3.26 \pm 0.15}$ & $23.90 \pm 0.45$ & $23.69 \pm 0.35$ & $23.93 \pm 0.37$ \\
 & Sparsity & $6.00 \pm 0.00$ & $\boldsymbol{2.0 \pm 0.0}$ & $6.00 \pm 0.00$ & $\boldsymbol{2.0 \pm 0.0}$ \\
 & Local stability ($r=1$) & $271.98 \pm 6.62$ & $167.79 \pm 3.88$ & $\boldsymbol{0.0 \pm 0.0}$ & $\boldsymbol{0.0 \pm 0.0}$ \\
\cline{1-6}
\bottomrule
\end{tabular}

\caption{{Optimization results. Numbers are mean property values of the explanation at each input, with 95\% confidence intervals. Rows are each property optimized explanation. Lower is more optimized. Bolded numbers are outside the CI of unbolded numbers. For stability, $r$ refers to the radius parameter from \cref{eq: stability} For faithfulness, $p$ refers to the distribution centered at the input from \cref{eq: fidelity}.}} 
\label{tab: optimization_results}
\end{table}

\subsection{Train and test points selection}
\label{appendix: train-and-test-selection}
\paragraph{Selection of training points.}
The training data was formed from a mix of pedagogical and informative points. As in the exploration phase, the informative training points are near the AI function's decision boundary (if the AI function is regression, we select these points randomly). The pedagogical points inform the user about a concept important to the task. For example, in forbidden features task, we may include one example where the attribution at the forbidden feature is $0$ (so the user should recognize the correct decision is ``forbidden feature \textit{was not} used'') and another example where everything else is the same, but the attribution is non-zero (so the user should recognize the correct decision is ``forbidden feature \textit{was} used'').

\paragraph{Selection of test points.}
For each small-scale synthetic study, we selected $30$ out of $500$ possible test points.

Recall that from the exploration phase, we have records of the proxy user performance on all $500$ test points across $10$ trials. This means that for each test point, we have $10$ observations on whether the proxy human answered that correctly or not for each explanation type. We can look at how the proxy user performed for each of these points to decide which ones to carry onto the small-scale synthetic study. 

\begin{itemize}
    \item \textit{$10$ instances where the proxy user performed the same, regardless of whether the explanation was robust, sparse, or faithful.} This protects against cases where real users would have shown differences across explanation types even though the proxies did not. 
    \item \textit{$10$ instances where the proxy user was correct when presented with the ``best explanation.''} This lets us test whether the real users perform better under the explanation type as expected. 
    \item \textit{$10$ instances where the proxy user was incorrect when presented with the ``best explanation.''} This protects against cases where real users would have performed well with the ``best explanations'' even though the proxy did not.
\end{itemize}
Above, the ``best explanation'' is the explanation type that resulted in the highest task-performance during the exploration phase.

\section{Qualitative codes}
\label{appendix: codes}
\begin{table}[h]
    \centering
    \begin{tabular}{l l p{3.5in} l}
         \textbf{Code} & \textbf{Grouping} & \textbf{Definition and Example} &  \textbf{Count} \\
         \toprule
         Weighted sum & Reasonable & Participant multiplied each attribution by the inputs, took the sum, and checked whether the sum was greater than a threshold & $\nicefrac{15}{156}$\\
         & \textit{Forward sim.}& \participant{If the sum of the measurements times what the alien researchers say each measurement affect’s the alien’s health exceeds 0 then the alien is healthy. Otherwise, if the sum is less than or equal to 0 then they’re sick.} & \\ \hline
         
         Threshold at 50 & Reasonable & Participant looked at inputs with negative attribution, checked whether greater than 50& $\nicefrac{4}{156}$\\
         & \textit{Forward sim.} & \participant{First thing im going to do is look at which of the aliens measurements are negative. I noticed that if a measurement is 50 or more over the negative number than the alien is sick.} &\\\hline
         
         Glow attribution 0 & Reasonable & Participant checked if the attribution for the ``glow'' feature was zero. & $\nicefrac{9}{156}$\\
         & \textit{Forbidden features}& \participant{My strategy for checking for compliance is to check whether the alien researcher has anything other than 0 under Glow, which is forbidden and should not be an influence in the diagnosis but can be measured.} & \\\hline

         Attribution vs change & Reasonable & Participant multiplied the sign of the changing input by the sign of the attribution& $\nicefrac{8}{156}$\\
         & \textit{Counterfactual sim.} & \participant{If measurement went up and researcher analysis is positive, risk goes up and vice versa. If measurement went up and researcher analysis is negative, risk goes down, and vice versa } & \\\hline

         Match explanation & Misuse & Participants tried to match the values of the inputs to the attributions & $\nicefrac{4}{156}$\\
         & & \participant{See if the measurements for the important categories match what the helpful information says. } & \\\hline

         Glow attribution & Misuse & Participants compared all other attributions to the glow (forbidden feature) attribution. & $\nicefrac{4}{156}$\\
         & & \participant{If the absolute value of the glow score is within a top 3 measurement meaning a greater deal of influence has been put on it, the case is non compliant} & \\\hline

         Extreme inputs & Ignore & Participants based their decision by looking at the extreme inputs (highest or lowest) & $\nicefrac{9}{156}$\\
         & & \participant{I will be focusing on the low scores of the measurement of the alien} & \\\hline

         Compare glow input & Ignore & Participants compared other input values to glow (forbidden feature) & $\nicefrac{4}{156}$\\
         & & \participant{Look at brainwave activity and core teperature scores and see how they compare to glow} & \\\hline

         Ideal ranges & Ignore & Participants formed a set of ``ideal input ranges'' in their head and based their decision on whether the inputs were within the ranges.& $\nicefrac{11}{156}$\\
         & & \participant{Glow is typically between 10-60; outside of this the alien is unhealthy.} & \\\hline

         Logic rules & Ignore & Participants formed a set of logic rules specific to the training examples. & $\nicefrac{16}{156}$\\
         & & \participant{If their core temperature is 40 or above, their glow level needs to be 40 and below} & \\\hline

         Explanation unhelpful & Ignore & Participant noted that the explanation was not helpful. & $\nicefrac{4}{156}$\\
         & & \participant{There seems to be a lot of cases where the researcher's estimates aren't even relevant, so I think I just have to guess on those. } & \\\hline

         No strategy & No strategy & Participant did not find a strategy.& $\nicefrac{16}{156}$\\
         & & \participant{I can't figure out a pattern.} & \\

         \bottomrule
          
    \end{tabular}
    \caption{Codes for participant decision-making strategies reported after training. Codes were grouped by: \textit{reasonable} for the task, \textit{misused} the explanations, \textit{ignored} the explanations, or had \textit{no strategy}. }
    \label{tab: codes}
\end{table}

\section{User Study Materials}

\subsection{End-of-study survey}
\label{appendix: end-survey}
We asked participants to fill out a short end-of-study survey. It was a series of seven-option likert questions, from -3 (strongly disagree) to 3 (strongly agree). Some of these questions were manipulation checks, and others were for post-study exploratory analyses: 
\begin{itemize}
    \item ``Information from the alien researcher is easy to use.'' Measures ease of use. 
    \item ``Information from the alien researcher narrows the number of measurements I have to look at.'' Measures sparsity. 
    \item ``Information from the alien researcher does not change much.'' Measures robustness. 
    \item ``Information from the alien researcher provides accurate information about how measurements affect alien health.'' Measures faithfulness. 
    \item ``Information from the alien researcher is helpful.'' Measures helpfulness.
    \item  ``I enjoy math problems.'' 
    \item ``I enjoy solving logic puzzles.''
\end{itemize}

\subsection{Manipulation check results}
\begin{figure}
    \centering
    \includegraphics[width=0.5\linewidth]{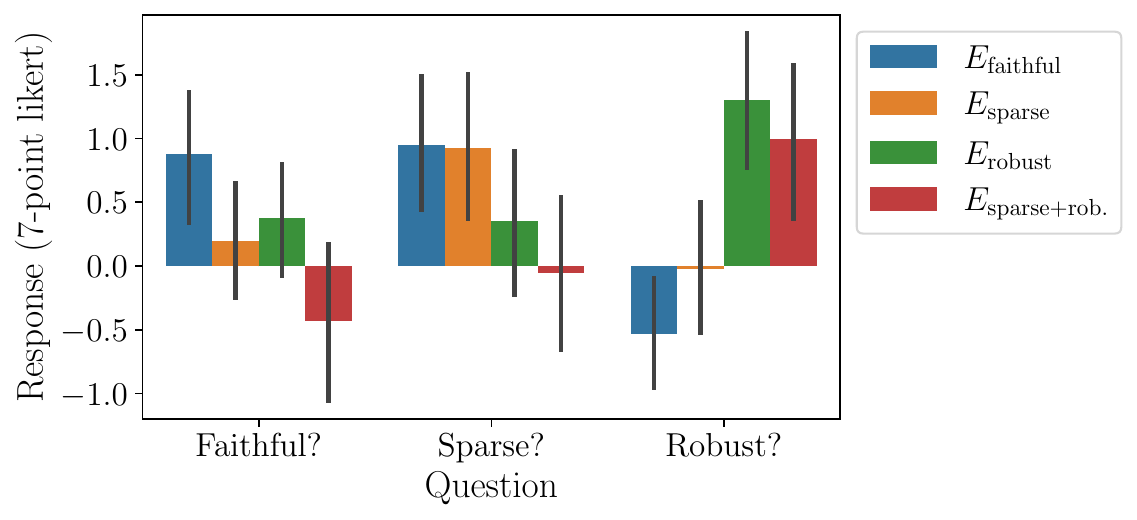}
    \caption{Responses to likert questions on the perceived properties of our explanations. X-axis is the likert question. Y-axis is the response, which ranges from -3 (strongly disagree) to +3 (strongly agree). Colors are explanation conditions. }
    \label{fig: manipulation-checks}
\end{figure}

In Figure \ref{fig: manipulation-checks}, we confirm that our participants perceive the explanations with different properties differently. 
The results are answers to the likert scale questions we presented in our end-of-study survey in \cref{appendix: end-survey}. 
Participants perceive explanations as we expected. For example, faithful explanations scored highest on the likert question that evaluated perceived faithfulness; this trend held for the other two properties. One unexpected result was that faithful explanations scored highly on perceived sparsity. This may be because of the wording of the sparsity question (``Information from the alien researcher narrows the number of measurements I have to look at.''); participants may have interpreted this as asking about general explanation usefulness. 

\subsection{Screenshots of study UI}
\label{appendix: screenshots}
\begin{figure}[h]
    \includegraphics[width=0.8\linewidth]{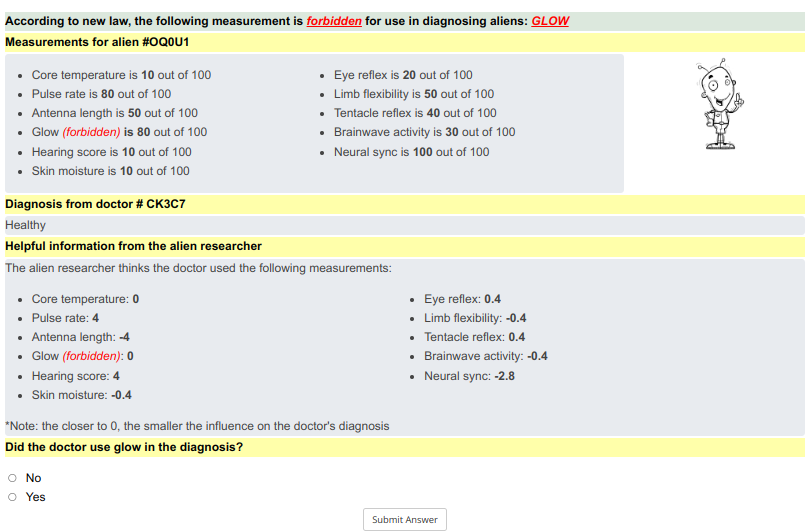} 
    \caption{UI for forbidden features}
    \label{fig: forbidden-features-ui}
\end{figure}

In \cref{fig: forbidden-features-ui}, the first block denotes the forbidden feature. The second block describes the alien's measurements, corresponding to the inputs $\mathbf{x}$. The third block is the doctor's diagnosis for the alien, corresponding to $\hat y(\mathbf{x})$. 
The fourth block contains the feature-attributions $\mathbf{w}$. The fifth block lets the user provide their decision on whether the forbidden feature was used in diagnosis, which corresponds to $h(\mathbf{x}_h)$. 

\begin{figure}[h]
    \includegraphics[width=0.8\linewidth]{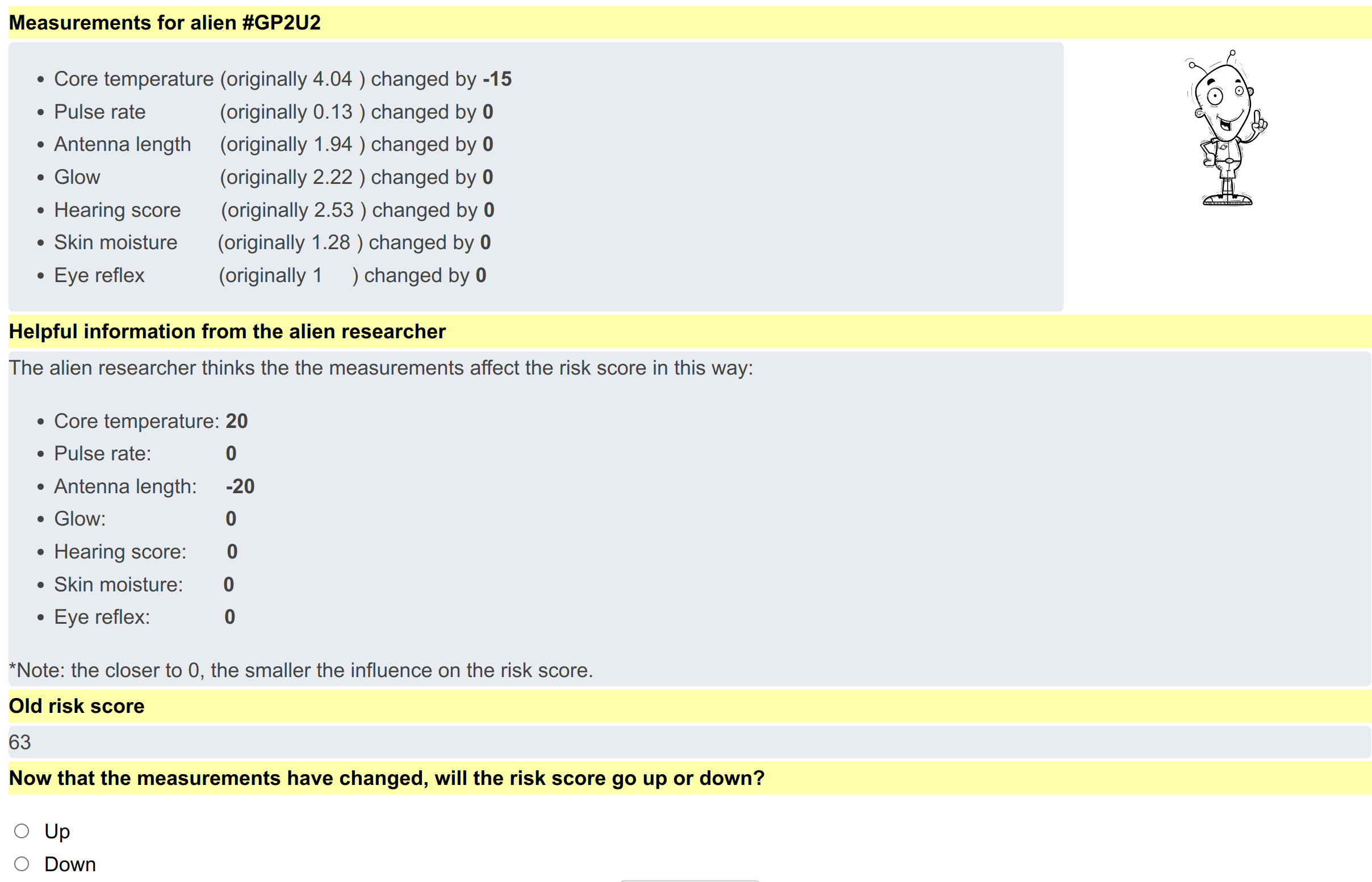} 
    \caption{UI for counterfactual simulation}
    \label{fig: counterfactual-simulation-ui}
\end{figure}

In \cref{fig: counterfactual-simulation-ui}, the first block denotes the alien's measurements and how they have changed since intake, corresponding to the inputs $\mathbf{x}$ and changes $\Delta_x$, respectively. The second block contains the feature-attributions $\mathbf{w}$. The third block contains the original risk score, which corresponds to $f(\mathbf{x})$. The final block is space for the user's decision on whether or not the risk score will go up or down.

\begin{figure}[h]
    \includegraphics[width=0.8\linewidth]{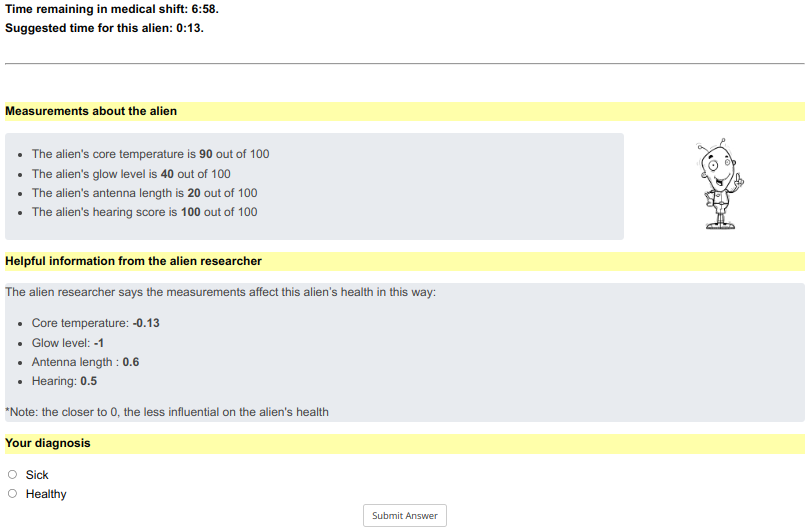}    
    \caption{UI for forward prediction}
    \label{fig: forward-prediction-ui}
\end{figure}

In \cref{fig: forward-prediction-ui}, the first block describes the alien's measurements, corresponding to the inputs $\mathbf{x}$. The second block contains the explanations $E(f, \mathbf{x})$. The third block lets the user provide their decision on the diagnosis, which corresponds to $h(\mathbf{x}_h)$.


\section{Exploration of alternate human proxy models}
\label{appendix: human-model-options}
\begin{figure}
    \centering
    \begin{subfigure}{0.33\linewidth}
        \includegraphics[width=1\linewidth]{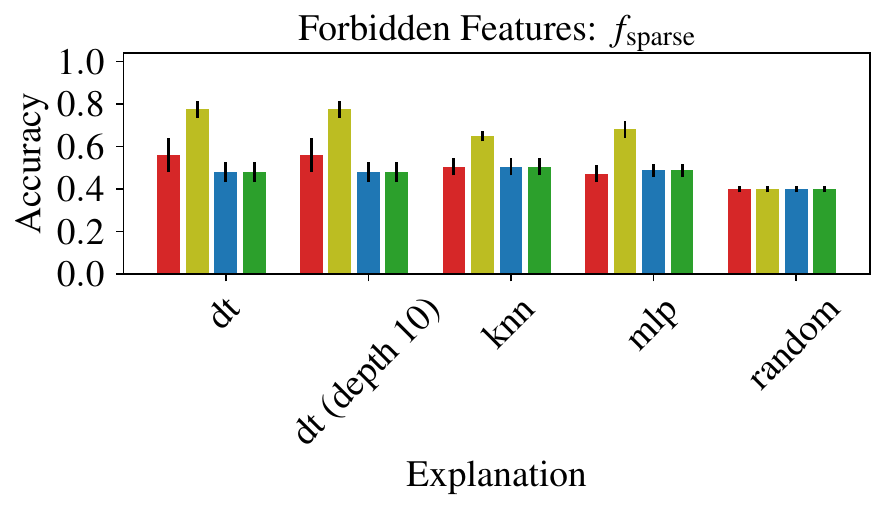}
    \end{subfigure}%
    \begin{subfigure}{0.33\linewidth}
        \includegraphics[width=1\linewidth]{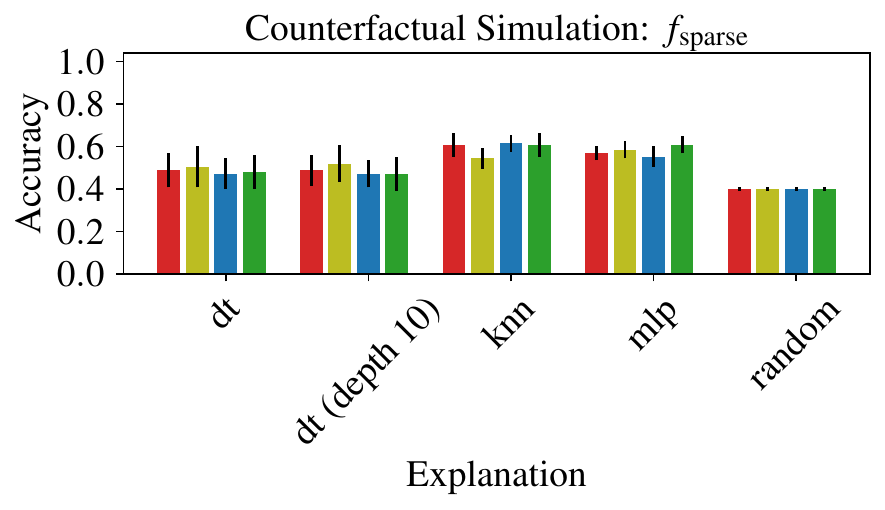}
    \end{subfigure}%
    \begin{subfigure}{0.33\linewidth}
        \includegraphics[width=1\linewidth]{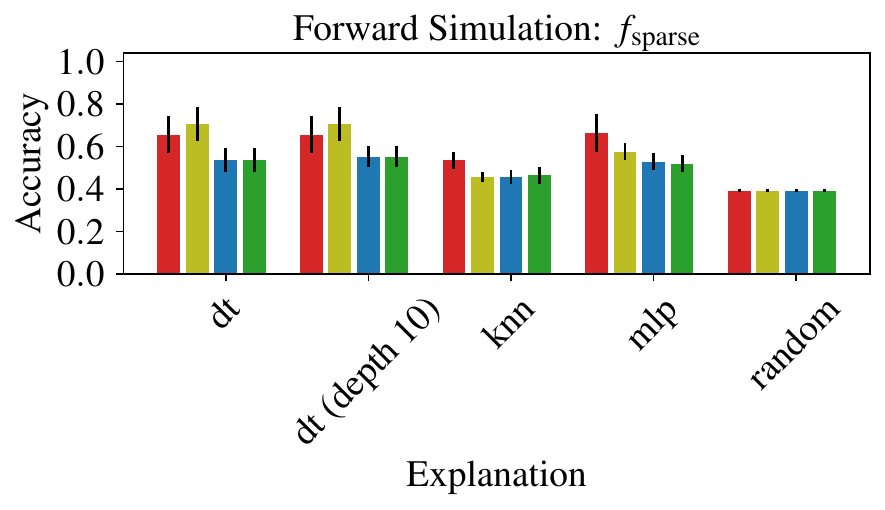}
    \end{subfigure}
    
    \begin{subfigure}{0.33\linewidth}
        \includegraphics[width=1\linewidth]{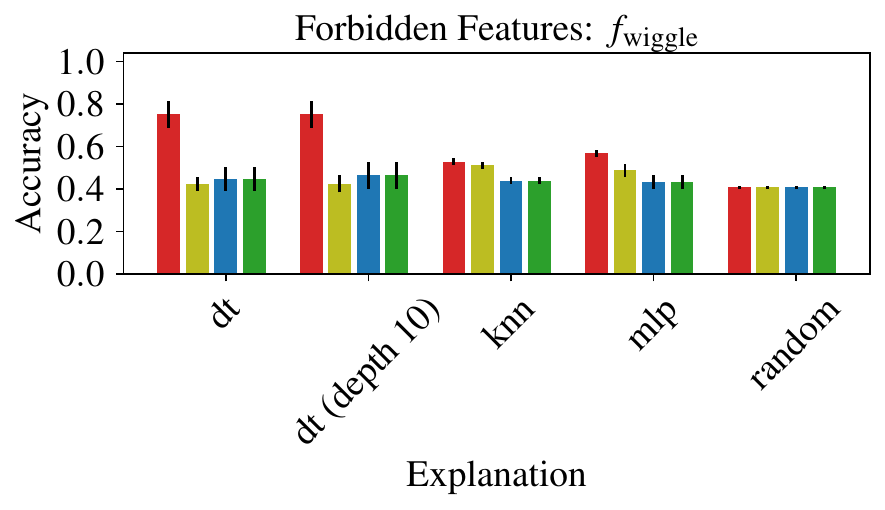}
    \end{subfigure}%
    \begin{subfigure}{0.33\linewidth}
        \includegraphics[width=1\linewidth]{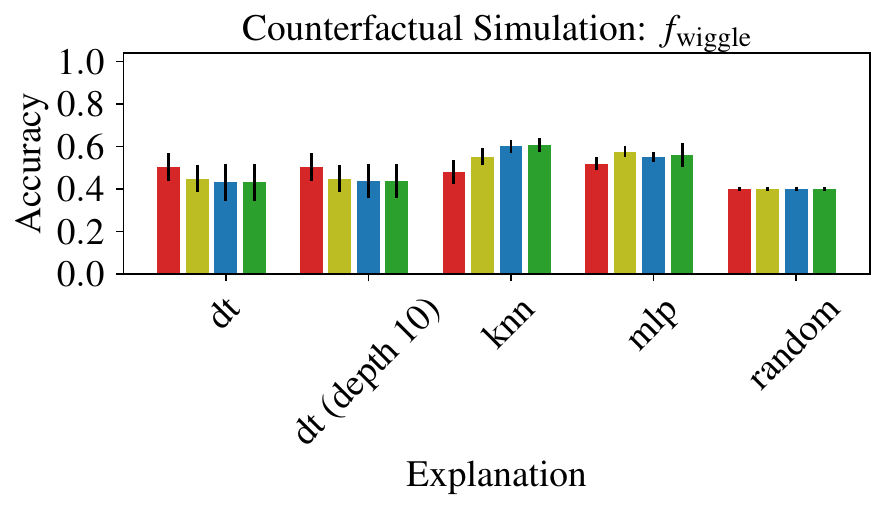}
    \end{subfigure}%
    \begin{subfigure}{0.33\linewidth}
        \includegraphics[width=1\linewidth]{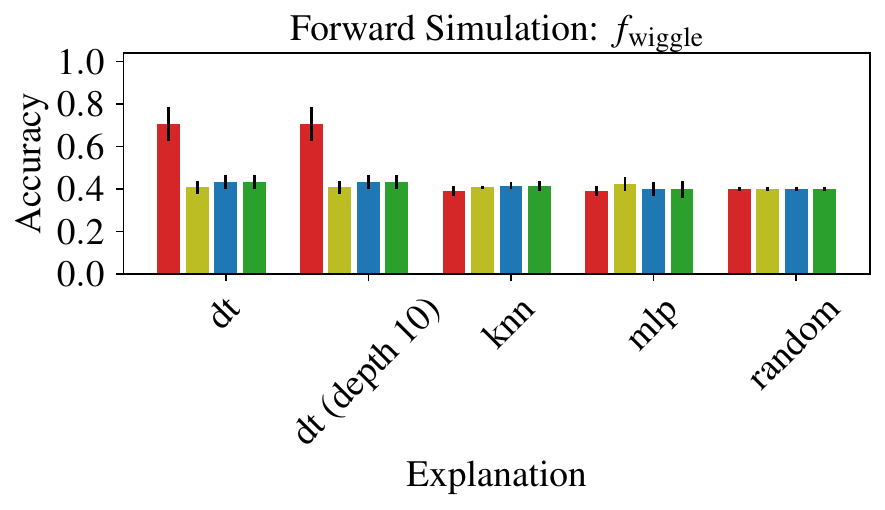}
    \end{subfigure}
    
    \begin{subfigure}{0.33\linewidth}
        \includegraphics[width=1\linewidth]{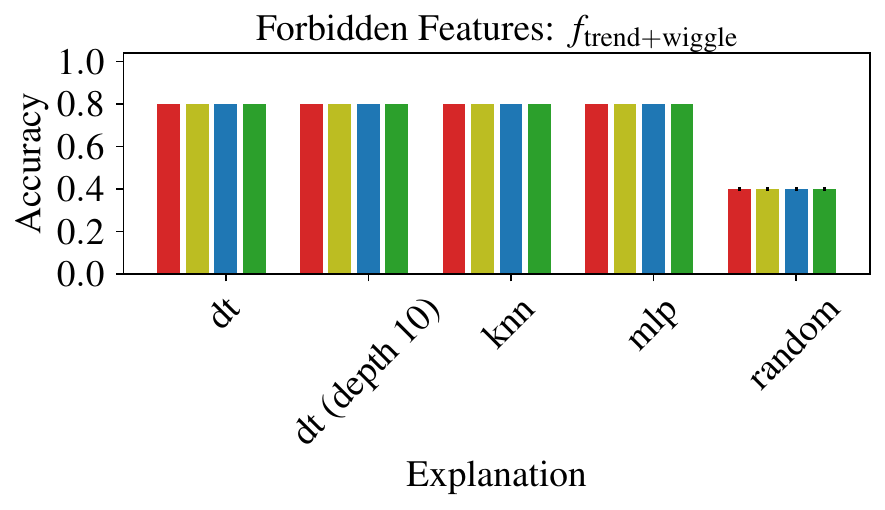}
    \end{subfigure}%
    \begin{subfigure}{0.33\linewidth}
        \includegraphics[width=1\linewidth]{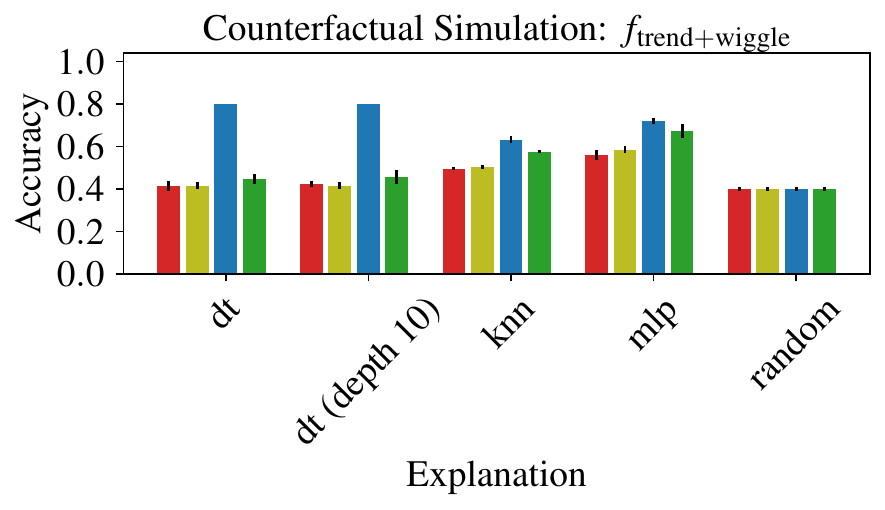}
    \end{subfigure}%
    \begin{subfigure}{0.33\linewidth}
        \includegraphics[width=1\linewidth]{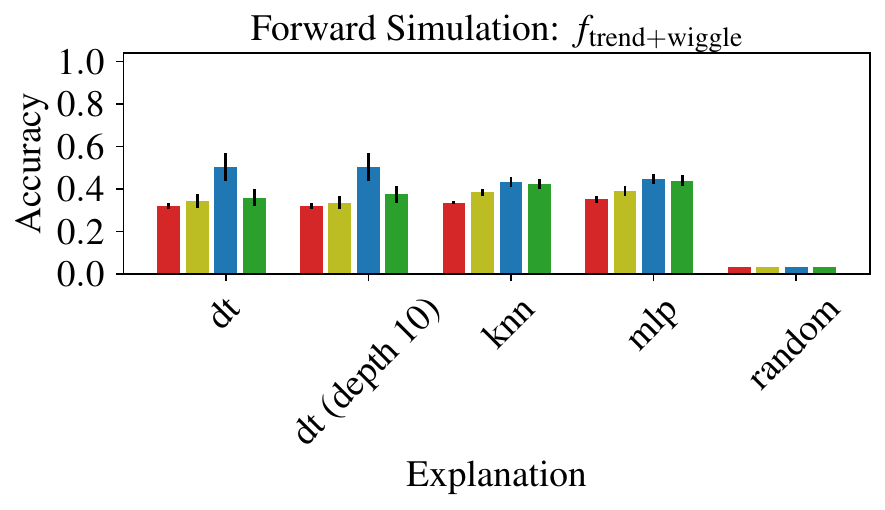}
    \end{subfigure}    
    \caption{Synthetic study results with different proxy model classes; DT and MLP mostly agree on ordering of explanations. Columns are tasks. Rows are AI functions. Each plot shows the performance of different explanation types, grouped by proxy model. }
    \label{fig: alternate-proxy-models}
\end{figure}


\end{document}